\documentclass[twocolumn]{aastex62}

\usepackage{xspace}

\newcommand{\coo}{CO$_2$}
\newcommand{\water}{H$_2$O}
\newcommand{\methane}{CH$_4$}

\newcommand{\update}{}

\newcommand{\tess}{{\em TESS}}
\newcommand{\spitzer}{{\em Spitzer}}
\newcommand{\gj}{GJ~1252}
\newcommand{\gjb}{GJ~1252b}

\newcommand{\rp}{1.180} 
\newcommand{\urp}{$\pm$0.078}
\newcommand{\vesc}{5.4}
\newcommand{\uvesc}{$\pm$0.8}

\newcommand{\tday}{1410}
\newcommand{\utday}{$^{+91}_{-125}$}
\newcommand{\depth}{149}
\newcommand{\udepth}{$^{+25}_{-32}$}
\newcommand{\te}{2458668.3575}
\newcommand{\ute}{$^{+0.0019}_{-0.0007}$}
\newcommand{\dt}{+1.4}
\newcommand{\udt}{$^{+2.8}_{-1.0}$}
\newcommand{\abUL}{0.41} 
\newcommand{\fLL}{0.40} 
\newcommand{\fabLL}{0.37} 
\newcommand{\ecoswnosign}{0.0025}
\newcommand{\ecosw}{+0.0025}
\newcommand{\uecosw}{$^{+0.0049}_{-0.0018}$}
\newcommand{\age}{3.9}
\newcommand{\uage}{0.4}

\accepted{August 10, 2022}

\shorttitle{No atmosphere on GJ~1252b}
\shortauthors{Crossfield et al.}

\begin{document}

\title{\gjb: A Hot Terrestrial Super-Earth With No Atmosphere }

\correspondingauthor{Ian J.\ M.\ Crossfield}
\email{ianc@ku.edu}

\author{Ian J.\ M.\ Crossfield}
\affiliation{Department of Physics and Astronomy, University of
  Kansas, Lawrence, KS, USA}

\author{Matej Malik} 
\affiliation{Department of Astronomy, University of Maryland, College Park, MD 20742, USA}

\author[0000-0002-0139-4756]{Michelle L. Hill} 
\affiliation{Department of Earth and Planetary Sciences, University of California Riverside, 900 University Ave, Riverside, CA 92521, USA}

\author{Stephen R.\ Kane} 
\affiliation{Department of Earth and Planetary Sciences, University of California Riverside, 900 University Ave, Riverside, CA 92521, USA}

\author{Bradford Foley} 
\affiliation{Earth and Planets Laboratory, Carnegie Institution for Science,
Washington, DC 20015, USA}

\author[0000-0001-7047-8681]{Alex S.\ Polanski} 
\affiliation{Department of Physics and Astronomy, University of
  Kansas, Lawrence, KS, USA}

\author{David Coria} 
\affiliation{Department of Physics and Astronomy, University of
  Kansas, Lawrence, KS, USA}

\author[0000-0002-2072-6541]{Jonathan Brande} 
\affiliation{Department of Physics and Astronomy, University of
  Kansas, Lawrence, KS, USA}

\author{Yanzhe Zhang} 
\affiliation{Department of Physics and Astronomy, University of
  Kansas, Lawrence, KS, USA}

\author{Katherine Wienke}  
\affiliation{Department of Physics and Astronomy, University of
  Kansas, Lawrence, KS, USA}

\author[0000-0003-0514-1147]{Laura Kreidberg} 
\affiliation{Max-Planck Institut f\"ur Astronomie, K\"onigstuhl 17, 69117, Heidelberg, Germany}

\author[0000-0001-6129-5699]{Nicolas B. Cowan} 
\affiliation{Department of Physics, McGill University, Montr\'{e}al,
QC H3A 2T8, Canada}
\affiliation{Department of Earth \& Planetary Sciences, McGill
University, Montr\'{e}al, QC H3A 2T8, Canada}

\author{Diana Dragomir} 
\affiliation{Department of Physics and Astronomy, University of New Mexico, Albuquerque, NM, USA}

\author{Varoujan Gorjian} 
\affiliation{Jet Propulsion Laboratory, California Institute of Technology, Pasadena, CA, USA}

\author{Thomas Mikal-Evans} 
\affiliation{Max-Planck Institut f\"ur Astronomie, K\"onigstuhl 17, 69117, Heidelberg, Germany}

\author{Bj\"orn Benneke} 
\affiliation{Departement de Physique, and Institute for Research on Exoplanets, Universite de Montreal, Montreal, Canada}

\author{Jessie L.\ Christiansen} 
\affiliation{Caltech/IPAC-NASA Exoplanet Science Institute, Pasadena, CA 91125, USA}

\author{Drake Deming} 
\affiliation{Department of Astronomy, University of Maryland, College Park, MD 20742, USA}


 \author{Farisa Y.\ Morales} 
 \affiliation{Jet Propulsion Laboratory, California Institute of Technology, Pasadena, CA, USA}

\begin{abstract}
In recent years the discovery of increasing numbers of rocky,
terrestrial exoplanets orbiting nearby stars has drawn increased
attention to the possibility of studying these planets' atmospheric
and surface properties. This is especially true for planets orbiting M
dwarfs, whose properties can best be studied with existing
observatories. In particular, the minerological composition of these
planets and the extent to which they can retain their atmospheres in
the face of intense stellar irradiation both remain unresolved.
Here we report the detection of the secondary eclipse of the
terrestrial exoplanet \gjb, obtained via ten eclipse observations
using the \spitzer\ Space Telescope's IRAC2 4.5\,\micron\ channel. We
measure an eclipse depth of \depth\udepth\ ppm, corresponding to a
day-side brightness temperature of \tday\utday~K. This measurement is
consistent with the prediction for a bare rock surface.  Comparing the
eclipse measurement to a large suite of simulated planetary spectra
indicates that \gjb\ has a surface pressure of $\lesssim 10$~bar ---
i.e., substantially thinner than the atmosphere of Venus. Assuming
energy-limited escape, even a 100~bar atmosphere would be lost in
$<$1~Myr, far shorter than our gyrochronological age estimate of
\age$\pm$\uage~Gyr. The expected mass loss could be overcome by mantle
outgassing, but only if the mantle's carbon content were $>$7\% by
mass --- over two orders of magnitude greater than that found in
Earth.  We therefore conclude that \gjb\ has no significant
atmosphere.  Model spectra with granitoid or feldspathic surface
composition, but with no atmosphere, are disfavored at $>$2$\sigma$.
The eclipse occurs just \dt\udt~min after orbital phase 0.5,
indicating $e \cos \omega$=\ecosw\uecosw, consistent with
a circular orbit. Tidal heating is therefore likely to be negligible to \gjb's
global energy budget.  Finally, we also analyze additional,
unpublished \tess\ transit photometry of \gjb\ which improves the
precision of the transit ephemeris by a factor of ten, provides a more
precise planetary radius of \rp\urp~$R_\oplus$, and rules out any
transit timing variations with amplitudes $\gtrsim 1$~min.

\vspace{1in}
\end{abstract}


\section{Introduction}
\label{sec:intro}

Rocky planets on short-period orbits are among the most common
planetary bodies known to emerge from the process of star and planet
formation \citep[e.g.,][]{fulton:2018}.  Though too small to retain a
primordial hydrogen envelope, such planets may produce secondary
atmospheres later in their evolution.  For example, in the Solar
System, rocky bodies exhibit a wide diversity of atmospheric surface
pressures from Venus (92~bar) to Earth and Titan ($\sim$1 bar) to Mars
(6~mbar) to Mercury and the Moon (negligible atmospheres).

The conditions under which terrestrial planets can retain sizable
atmospheres under different irradiation levels, timescales, types of
host star, and planet masses, radii, and surface gravity all remain
areas of active research.  While an exoplanet's atmosphere can be
studied via transit and/or eclipse observations, transmission
spectroscopy has so far failed to conclusively determine the
properties (or absence of) any rocky planet's atmosphere \citep[e.g.,
  see][]{wordsworth:2022}. To date, emission measurements have offered
the best prospects for studying the properties of terrestrial
exoplanets.

Secondary eclipses of several rocky planets were detected at optical
wavelengths by the {\em Kepler/K2} missions
\citep[e.g.,][]{batalha:2011,sheets:2014,malavolta:2018}. Unfortunately,
such measurements often suffer from a degeneracy: optical eclipses
represent a combination of reflected/scattered light and thermal
emission, with no empirical way to determine the relative
contributions of each.

Until now thermal infrared radiation has been measured from only two
terrestrial exoplanets, LHS~3844b \citep{kreidberg:2019} and K2-141b
\citep{zieba:2022}.  \spitzer\ 4.5\,\micron\ observations of these
planets' eclipse and phase curves revealed no phase offset and
suggested an upper limit to the atmospheric surface pressure; for
example, the data set for LHS~3844b indicates
$P_\mathrm{surf}\lesssim 10$~bar.  

In this paper we report 4.5\,\micron\ eclipse photometry that reveals
thermal emission, and similar constraints on the atmosphere, of \gjb,
the smallest exoplanet for which such a measurement has been made to
date.  The planet has an Emission Spectroscopy Metric
\citep[ESM;][]{kempton:2018} of 17, slightly larger than that of
K2-141b and a factor of two smaller than LHS~3844b. \gjb\ was
identified by the \tess\ project as \tess\ Object of Interest (TOI)
1078.01 in data from Sector 13, the last southern sector to be
observed in the first year of \tess\ operations. \cite{shporer:2020}
confirmed the planetary nature of the signal using a combination of
\tess\ photometry and HARPS radial velocities. They reported a planet
orbiting an M3V star with radius of $1.193 \pm 0.074\, R_\oplus$,
{\update and the planet's mass is $1.32 \pm 0.28\, M_\oplus$ (Luque et
  al., in review).}

In Sec.~\ref{sec:obs} we present our \tess\ and {\em Spitzer}
observations and our analyses of these data. Sec.~\ref{sec:interp}
then discusses these measurements in light of a set of models of
planetary spectra, leading us to conclude that any atmosphere on
\gjb\ likely has a surface pressure of $\lesssim$10~bar.
Sec.~\ref{sec:escape} presents our predictions for atmospheric escape
from \gjb, which leads us to conclude that even an atmosphere with
surface pressure $>$10~bar would be lost on a timescale much shorter
than the system age. Finally, we close with a discussion of \gjb\ in
the context of similar measurements of other rocky exoplanets in
Sec.~\ref{sec:conclusions}.

\section{Observations}
\label{sec:obs}

\subsection{New \tess\ Transit Photometry}

Subsequent to the mid-2019 \tess\ Sector~13 photometry used to first
discover \gjb\ \citep{shporer:2020}, the system was
re-observed during the \tess\ Sector 27 Campaign using Camera 2 from
2020 July 5 to 2020 July 30.  In this section we describe our combined
analysis of both the original Sector 13 and the new Sector 27 data. By
performing a global fit on data sets separated by nearly a year, we
further refine the orbital and planetary properties of \gjb.

We downloaded both Sector 13 and 27 Presearch Data Conditioning (PDC)
time series measurements from MAST. PDC-level data products are
corrected for instrumental systematics and contamination from nearby
stars. Our analysis used the \texttt{LightKurve} package
\citep{lightkurve:2018} to perform 5 iterations of $3\sigma$ outlier
rejection on data points above the median flux level. This removed
0.2\% of data from the lightcurve. To remove any remaining flux
variations we flattened the lightcurve using a Savitzky-Golay filter
\citep{savitzky:1964} {\update after first masking out the transits
  (with one transit duration on either side) before applying the
  filter to ensure the transit features are not affected.}

We fit the flattened lightcurve using the \texttt{exoplanet} package
\citep{foremanmackey:2021} which uses a Hamiltonian Monte Carlo (HMC)
routine to explore the posterior probability distribution. Assuming a
circular orbit with the period ($P$), time of inferior conjunction
($T_{0}$), scaled planet radius ($R_P/R_*$), impact parameter ($b$),
transit duration ($T_{14}$), and mean flux offset ($\mu$) as free
parameters we minimized a negative log-likelihood function. Our
analysis held quadratic limb-darkening coefficients constant at
$u_0$=0.2800 and $u_1$=0.3683 \citep[values taken
  from][]{claret:2018a}. We used the values obtained from the
minimizer as initial positions for 32 parallel chains and ran the HMC
for 2,000 tuning steps and 4,000 sampling steps per chain. Loose
Gaussian priors (much wider than the final posteriors) were placed on
$P$ {\update (mean 0.51 days, width $\sigma$~0.05~days)} and
$T_\mathrm{conj}$ {\update (mean 1668.0, width 0.1)} to prevent the
sampler from wandering too far astray. {\update The final Gelman-Rubin
statistic of the HMC runs were $<1.01$ for all parameters.}

\begin{figure}
\centering
\includegraphics[width=0.5\textwidth]{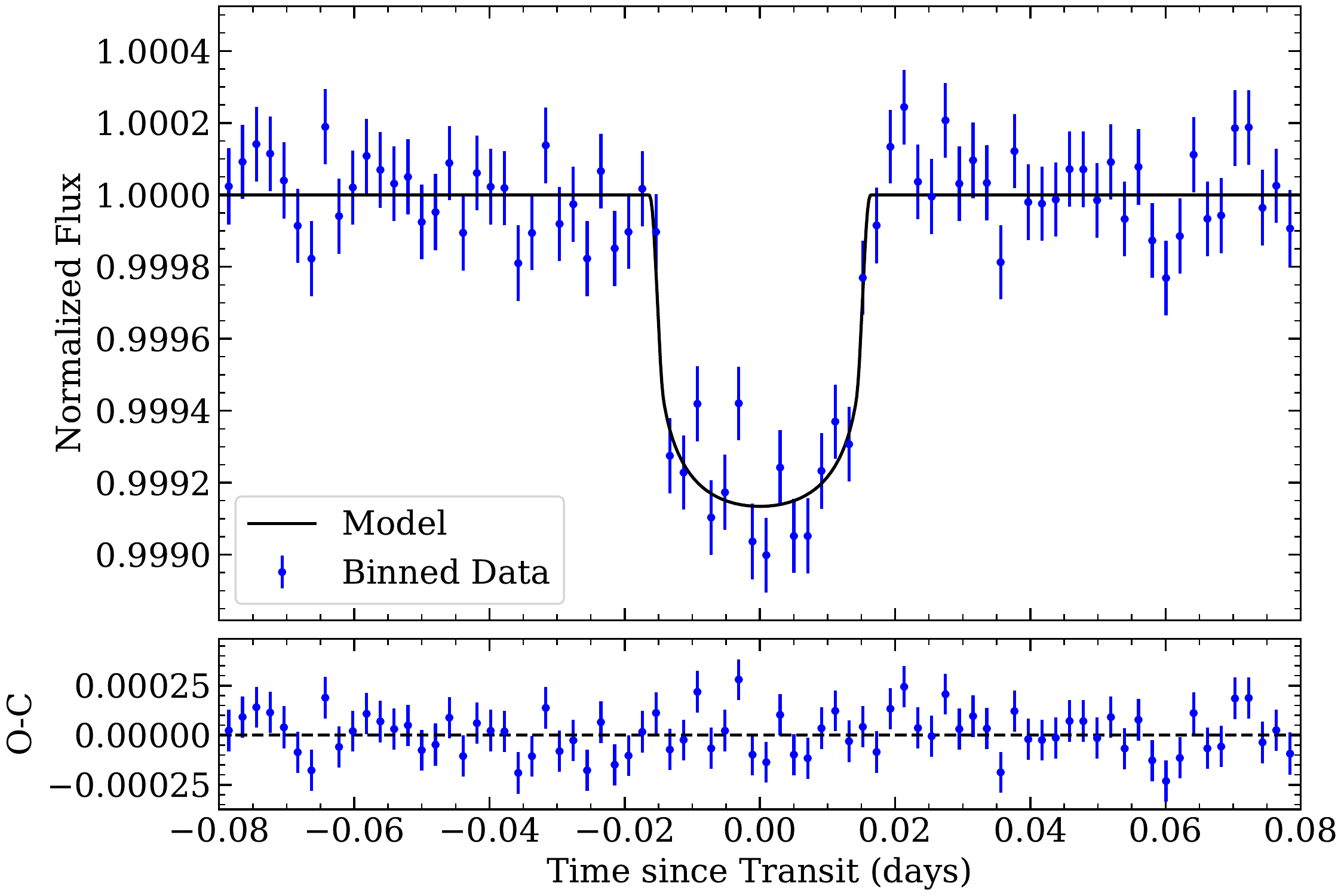} 
\caption{Folded and binned two-sector {\em TESS} lightcurve (blue points) with
  the median model plotted in black. Residuals (data minus model) are
  shown in the lower panel.}
\label{fig:tess_fold}
\end{figure}

The median values and their 1-$\sigma$ uncertainties are listed in
Table~\ref{tab:params}, and the model fit to binned \textit{TESS} data
is shown in Figure \ref{fig:tess_fold}. Our results agree well with
those of the original discovery paper \citep{shporer:2020}. As an
independent check, we also analyzed the full, two-sector \tess\ data
set using the transit light curve code described in numerous similar
{\em K2} studies \citep[e.g.,][]{crossfield:2015a,crossfield:2016} and
found consistent parameters in all cases.

{\update We analyze only the transits because the current {\em TESS}
  data cannot usefully constrain the planet's eclipse depth. In the
  {\em TESS} bandpass the contribution from either thermal emission
  (Fig.~\ref{fig:rockspec}) or reflected light
  (Sec.~\ref{sec:fromEtoB}) is $\lesssim$15~ppm, significantly smaller
  than our {\em TESS} transit depth precision of 61~ppm
  (Table~\ref{tab:params}).}

Finally, we also performed a search for transit-timing variations
(TTVs) in the TESS data, again using the \texttt{exoplanet} package.
If present, TTV signals could indicate the presence of undiscovered
companions due to mutual gravitational interactions or orbital decay
due to tidal effects. Although the S/N of individual transit events is
quite low, we find no evidence for TTVs with amplitudes
$\gtrsim1$~min, consistent with a lack of strongly perturbing
companions. The deviation of \gjb's individually-measured transit
times is consistent with a linear ephemeris across both sectors of
TESS data. With nearly a year separating these two Sectors of
\textit{TESS} data, our analysis reduces the uncertainty on the period
by an order of magnitude (see Table~\ref{tab:params}).




\begin{deluxetable*}{l l l l}[bt]
\tabletypesize{\scriptsize}
\tablecaption{  Planet Parameters \label{tab:params}}
\tablewidth{0pt}
\tablehead{
\colhead{Parameter} & \colhead{Units} & \colhead{Value} & \colhead{Source} 
}
\startdata
\multicolumn{4}{l}{\hspace{0.1in}\em Stellar parameters:}\\
$R_*$ &  $R_\odot$ & $0.391\pm 0.020$ & \cite{shporer:2020} \\
     $M_*$ &  $M_\odot$ & $0.381\pm 0.019$ & \cite{shporer:2020} \\
$T_\mathrm{eff}$ &  K & $3458 \pm 137$      & \cite{shporer:2020} \\
age & Gyr & \age$\pm$\uage & This work, derived \\
\multicolumn{4}{l}{\hspace{0.1in} TESS {\em Transit parameters:}}\\
$T_{0}$ & $BJD_{TDB}  $ & $2458668.09748\pm 0.00032$ & This work, fit \\
          $P$ &          d & $0.51824160\pm {0.00000069}$ & This work, fit\\
          $i$ &        deg & $84.8\pm 3.2$ & This work, derived\\
    $R_P/R_*$ &         -- & $0.0277\pm 0.0011$ & This work, fit\\
      $a/R_*$ &         -- & $5.03\pm 0.27$ & This work, derived\\  
     $T_{14}$ &         hr & $0.724 \pm 0.013$ & This work, fit\\
         $b$ &         -- & $0.42 \pm 0.24$ & This work, fit\\
        $a$ &         AU & $0.00915^{+0.00015}_{-0.00015}$ & This work, derived\\
 $S_{inc}$ &      $S_\oplus$ & $233^{+48}_{-41}$ & This work, derived\\
     $R_P$ &      $R_\oplus$ & \rp\urp & This work, derived\\
     $M_P$ &      $M_\oplus$ & $1.32 \pm 0.28$ & {\update Luque et al., in review}\\
\multicolumn{4}{l}{\hspace{0.1in}Spitzer {\em Eclipse parameters:}}\\
$T_E$ & $BJD_{TDB}$ & \te\ute & This work, fit \\
$\delta_{IRAC2}$ & ppm & \depth\udepth & This work, fit  \\
$dt$ & min & \dt\udt & This work, derived \\
$e \cos \omega$ & -- & \ecosw\uecosw & This work, derived \\
$T_\mathrm{day}$ & K & \tday\utday & This work, derived \\
$f $ & -- & $>$\fLL\ & This work, derived\\
$A_B$ & -- & $<$\abUL\ & This work, derived\\
$f \left(1 - A_B \right)$ & -- & $>$\fabLL\ & This work, derived \\
\enddata
\tablenotetext{}{Two-sided intervals indicate  68.3\% (1$\sigma$) confidence; one-sided intervals indicate 95.4\% (2$\sigma$) confidence.}
\end{deluxetable*}

\begin{deluxetable}{rrrrrr}
\tabletypesize{\scriptsize}
\tablecaption{\emph{Spitzer}/IRAC 4.5\,\micron\ Observations  \label{tab:obs}}
\tablewidth{0pt}
\tablehead{\colhead{} & \colhead{Start Date} & \colhead{Start}  & \colhead{End} & \colhead{$\delta$ }& \colhead{$\sigma_\delta$ }  \\ 
\colhead{AOR} & \colhead{[$BJD_{TDB}$]} & \colhead{ Phase}  & \colhead{ Phase} & \colhead{ [ppm]}& \colhead{[ppm]} }

\startdata
 71407360 &  2458868.859948 & 0.3917 &  0.6228  & 86    & 88    \\
 71407872 &  2458867.834249 & 0.4125 &  0.6436  & 101   & 97   \\
 71408384 &  2458866.277380 & 0.4084 &  0.6395  & $-$4    & 88   \\
 71408640 &  2458864.190715 & 0.3819 &  0.6131  & 168   & 83   \\
 71408896 &  2458863.686588 & 0.4092 &  0.6403  & 44    & 87   \\
 71409152 &  2458863.163738 & 0.4003 &  0.6314  & 123   & 86   \\
 71409408 &  2458861.090252 & 0.3993 &  0.6304  & 221   & 83   \\
 71409664 &  2458860.062523 & 0.4162 &  0.6473  & 357   & 103  \\
 71409920 &  2458859.008577 & 0.3825 &  0.6136  & 307   & 88   \\
 71410176 &  2458869.902201 & 0.4028 &  0.6340  & 137   & 83   
\enddata
\end{deluxetable}

\subsection{Spitzer Eclipse Photometry and Analysis}
\subsubsection{Eclipse Observations}
Soon after the \tess\ project's announcement of a planet candidate
around \gj, and before the planet's confirmation by
\cite{shporer:2020}, we identified the planet candidate as a promising
target for thermal infrared emission measurements obtained during secondary
eclipse. Using preliminary information provided in the \tess\ alert and
the \tess\ Input Catalog \citep[TIC;][]{stassun:2017}, we estimated
that a coordinated campaign of \spitzer\ eclipse observations could
detect the planet's eclipses. We therefore scheduled ten 4.5\,\micron\ eclipse
observations as part of Spitzer Program 14084
\citep{crossfield:2018spitzer}.

We observed the ten eclipses of \gjb\ over ten days in January 2020. The final
observations were taken on UT 2020-01-21, less than ten days before
\spitzer\ was deactivated on 2020-01-30.  Each eclipse observation was
an identical, 2.9\,hr, continuous, staring observation centered on the
predicted time of secondary eclipse (i.e., orbital phase 0.5). The
visits consisted of 5120 subarray frames with 2\,s integrations, taken
with the IRAC2 4.5\,\micron\ camera \citep{fazio:2005}. The
observations used IRAC's peak-up mode to place the star near a
well-characterized and well-behaved region of the detector, in order
to minimize the effect of IRAC's well-known intrapixel sensitivity
variations. Table~\ref{tab:obs} lists the times
and orbital phases of each of the ten eclipse observations.

\subsubsection{Eclipse Analysis}
We analyzed the \spitzer\ photometry using Pixel-Level Decorrelation
\citep[PLD;][]{deming:2015}, which models the systematics-dominated
\spitzer\ light curve as a linear combination of basis vectors derived
from each pixel's time series. Specifically,  we use the formulation
\begin{equation}
  f_{ik} = \left( \sum_{j=1}^{N+M} c_{jk} v_{ijk}  \right) m_{ik}
\end{equation}
where $f_{ik}$ is the modeled flux at the $i^\mathrm{th}$ timestep of
the $k^\mathrm{th}$ eclipse visit, $c_{jk}$ is the scaling coefficient
for the corresponding basis vector $v_{ijk}$, and $m_{ik}$ is the
purely astrophysical model of a secondary eclipse. The basis vectors
$v_{ijk}$ always include the $N$ individual pixel time series from the
$k^\mathrm{th}$ visit (the ``pixel-level'' data essential to PLD) and
may also include $M$ low-order temporal trends (for which $v_{ijk} =
t_{ijk}^p$ for $p=0, 1, ... (M-1)$\,) or other systematic vectors
against which the data will be decorrelated. In our analysis we
included a linear trend with time in order to remove a slow, long-term
drift.  We parameterized the eclipse model ($m_{ik}$) using the
\cite{mandel:2002} formulae for the occultation of an object with
uniform surface brightness, with its only free parameters being the
time of mid-eclipse $T_E$ and the fractional eclipse depth $\delta$.

PLD is often applied by simultaneously sampling the posterior
distribution of the nuisance parameters $c_{jk}$ as well as the
astrophysical parameters of interest. In our case, this turned out to
be intractable. With ten eclipse visits, the use of $N=9$ pixels and
$M=1$ (constant scaling, linear polynomial trend) would require
marginalizing over 100 nuisance parameters to obtain measurements of
just two astrophysical parameters, $T_E$ and $\delta$.

Instead, we determine $T_E$ and $\delta$, and their uncertainties, as
follows. For each combination of these parameters, we calculate a
model eclipse light curve and divide the observed flux $F_{ik}$ by
it. The result is
\begin{equation}
  \frac{F_{ik}}{m_{ik}} \approx  \sum_{j=1}^{N+M} c_{jk} v_{ijk}   + \epsilon_{ik}
  \label{eq:lsq}
\end{equation}
where $\epsilon_{ik}$ represent the measurement errors.  We then
directly solve Eq.~\ref{eq:lsq} at each point for the $c_{jk}$ using
weighted linear least squares. This approach allows us to directly
sample the two-dimensional $(T_E, \delta)$ plane while also accounting
for the interrelationships of these astrophysical parameters on the
$c_{jk}$ nuisance parameters. This approach is similar in some ways to
the PLD analysis of \spitzer/IRAC microlensing observations
\citep{dang:2020}; the main difference is that that work had more than
two astrophysical parameters of interest and so used MCMC sampling
instead of directly calculating a grid of likelihood values.

We set the weight of each observation equal to $1/\sigma^2$, where
$\sigma$ is the 68.3\% central confidence interval on the dispersion
of the residuals to an initial fit. Flux measurements $F_{ik}$ are
set to zero weight if they deviate by more than $5\sigma$ from the
nominal model.

\begin{figure}[bt!]
\begin{center}
\includegraphics[width=3in]{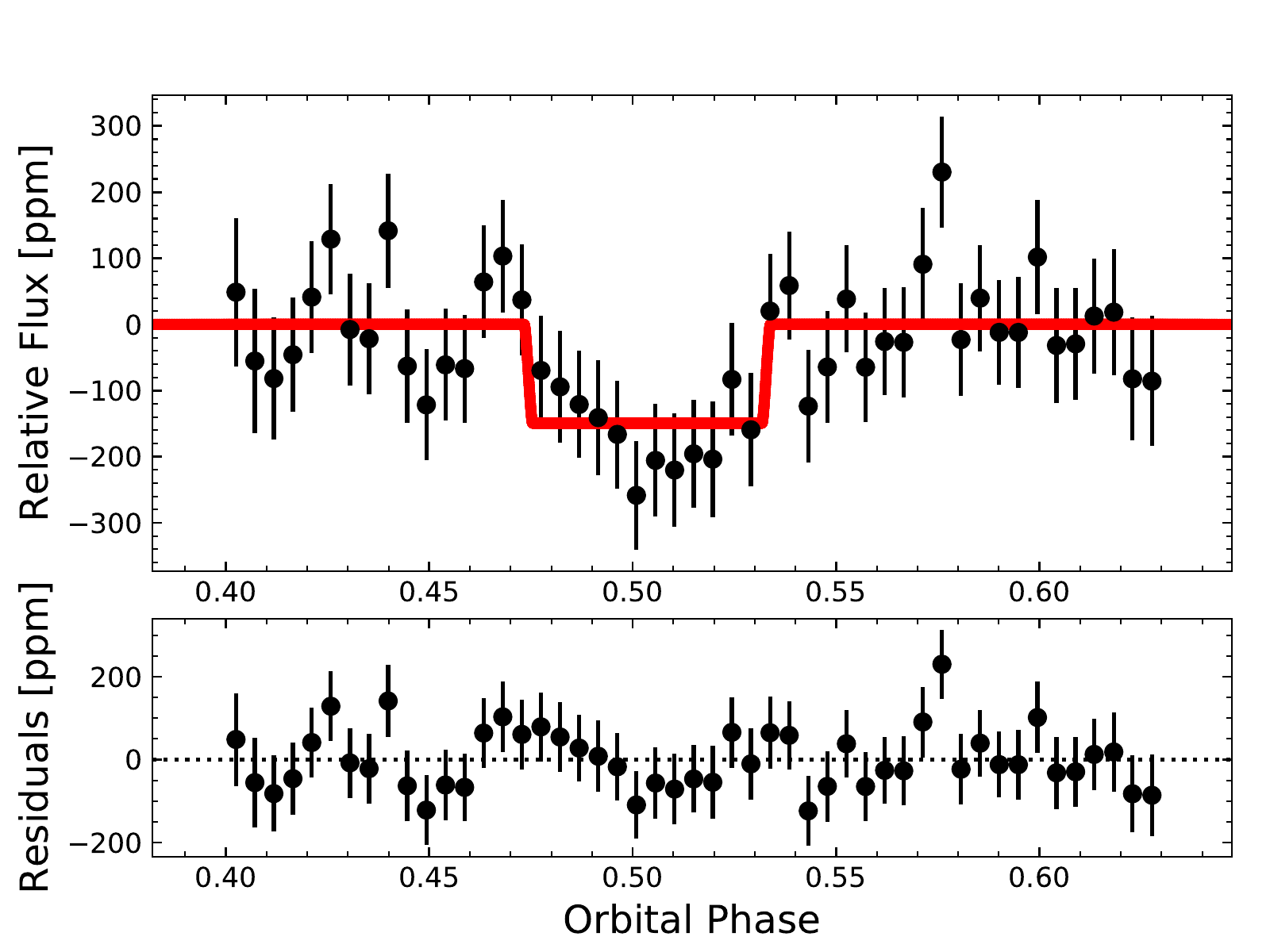}
\caption{\label{fig:eclipse} \spitzer\ 4.5\,\micron\ photometry and secondary
  eclipse fit ({\em top}) and residuals to the fit ({\em bottom}). The
  photometry is shown after detrending for systematic effects,
  combining the photometry from all ten eclipse visits, and binning
  down to a two-minute cadence. The measured depth is \depth\udepth~ppm. }
\end{center}
\end{figure}

The resulting stacked, detrended eclipse light curve shown in
Fig.~\ref{fig:eclipse} shows a clear flux decrement of
\depth\udepth~ppm at the expected time of eclipse.  Our measured
eclipse parameters are listed in Table~\ref{tab:params}, while the
light curves for each individual visit, as well as Allan deviation
plots of the residuals to each visit, are shown in the Appendix.
{\update As two checks on our measured eclipse depth, we also
  calculated the weighted mean of the eclipse depths from each
  individual visit, and also conducted a joint analysis in which the
  time of eclipse was held fixed to orbital phase 0.5.  The weighted
  mean is $150 \pm 28$~ppm, while the fixed-time analysis yields a
  depth of $140^{+35}_{-20}$~ppm; both these values are consistent
  with the value obtained from our joint fit.} 

\subsection{Stellar Age From Gyrochronology}
Finally, to interpret our measurements of \gjb's thermal emission and
estimate its atmospheric evolution (described below), we need to
estimate a stellar age.  Inferring ages for mature, field M dwarfs is
notoriously challenging; one promising avenue for all but the latest M
spectral subtypes is the use of gyrochronology.  Using the spindown
analysis of \cite{engle:2018} together with \gj's stellar rotation
period of $64\pm4$~d \citep{shporer:2020}, we estimate \gj's age to be
\age$\pm$\uage~Gyr. This indicates that \gj\ is somewhat younger than
LHS~3844 \citep[7.8$\pm$1.6~Gyr;][]{kane:2020}, an estimate broadly
consistent with the non-detection of stellar flares during the first
sector of \tess\ observations of \gj\ \citep{howard:2022}. This result
also indicates that the \gj\ system is somewhat {\update younger} than the $7.8
\pm 1.6$~Gyr LHS~3844b \citep{kane:2020}.

\section{Timing, Tides, and Atmospheres}
\label{sec:interp}
Our analysis of the Spitzer/IRAC2 4.5\,\micron\ photometry clearly detects the
secondary eclipse signal, {\update which has} a depth of
$\delta$=\depth\udepth~ppm and a timing offset from orbital phase
0.5 of just \dt\udt~min {\update (the joint posterior distribution of eclipse depth and eclipse timing is shown in the Appendix)}. Here we discuss the implications of these measurements: first of the eclipse timing in Sec.~\ref{sec:timing}, and then of the measured depth in Sec.~\ref{sec:depthresults}

\subsection{Eclipse Time and Implications}
\label{sec:timing}
The offset of the eclipse time from orbital phase 0.5 constrains the
combination of orbital parameters $e \cos \omega$ \citep{winn:2010};
for \gjb\ we find $e \cos \omega$=\ecosw\uecosw. This result is
consistent with zero at the 1.4$\sigma$ level, so we do not take this
measurement as evidence of an eccentric orbit. Regardless, the $e \cos
\omega$ measurement further justifies the assumption of low
eccentricity in the radial velocity analysis of \cite{shporer:2020}.
We conducted a reanalysis of their radial velocity data while
incorporating this new constraint on $e \cos \omega$, finding results
consistent with those of the discovery paper.

Although \gjb's orbit would quickly circularize in the absence of
other perturbers, small planets orbiting M dwarfs are often found in
multi-planet systems and additional bodies in the system could cause
\gjb\ to stay on an eccentric orbit.  Although no such bodies were
indicated by our TTV analysis, it is still possible that tidal heating
could act as an additional heat source in \gjb.  We estimate this
heating level following the prescription of \cite{henning:2009} and
assuming a Love number $k_2=0.3$ and tidal quality factor $Q=10^4$,
approximately appropriate for super-Earths
\citep{miguel:2011,millholland:2019}.  Further assuming that
$e=\ecoswnosign$, we find that tidal heating should contribute
      {\update only $4 \times 10^{16}$~W to \gjb's total energy
        budget, negligible \citep[unless tidal heating is enhanced via
          a significantly nonzero axial tilt;][]{millholland:2019}
        compared to the roughly $6 \times 10^{19}$~W of starlight
        absorbed by the planet}.



\subsection{Eclipse Depth and Implications}
\label{sec:depthresults}
Here we consider the implications of our eclipse measurement on the
surface and atmospheric properties of \gjb.  We first consider our
results in the context of global energy balance in
Sec.~\ref{sec:phenom}, and then in the context of a suite of one-dimensional
radiative transfer models in Sec.~\ref{sec:radtrans}.

\subsubsection{From Eclipse Depth to Brightness Temperature}
\label{sec:fromEtoB}
Converting an eclipse depth measurement into a brightness temperature
requires an estimate of both $R_P/R_*$ and of the stellar flux density
at the relevant wavelengths. Since M dwarfs such as \gj\ have emergent
spectra that differ considerably from simple blackbodies, appropriate
stellar spectra must be used. We used the BT-Settl suite of stellar
models \citep{allard:2014}, interpolating across the model grid using
the $T_\mathrm{eff}$, $\log_{10} g$, and [Fe/H] from
\cite{shporer:2020}.  The result is that \gjb's day side has a
4.5\,\micron\ brightness temperature of \tday\utday~K, considerably
hotter than the equilibrium temperature of $1089\pm69$~K reported by
\cite{shporer:2020}.

We note that reflection or scattered light contributes only
$\left( R_p/a \right)^2 A_g \approx \left(30\, \mathrm{ppm} \times A_g \right)$ to the eclipse depth, where $A_g$ is the
planet's 4.5\,\micron\ broadband geometric albedo. With $A_g <0.5$
expected for most typical minerals \citep{mansfield:2019} and $<0.1$
for lava or volcanic glasses \citep{essack:2020,modirrousta:2021}, the contribution of
surface reflection to our measurement is $<15$~ppm, smaller than our
{\em Spitzer} (or \tess) measurement precision.


\subsubsection{Energy Balance and Atmospheric Circulation}
\label{sec:phenom}
\begin{figure}[bt!]
\begin{center}
\includegraphics[width=3.4in]{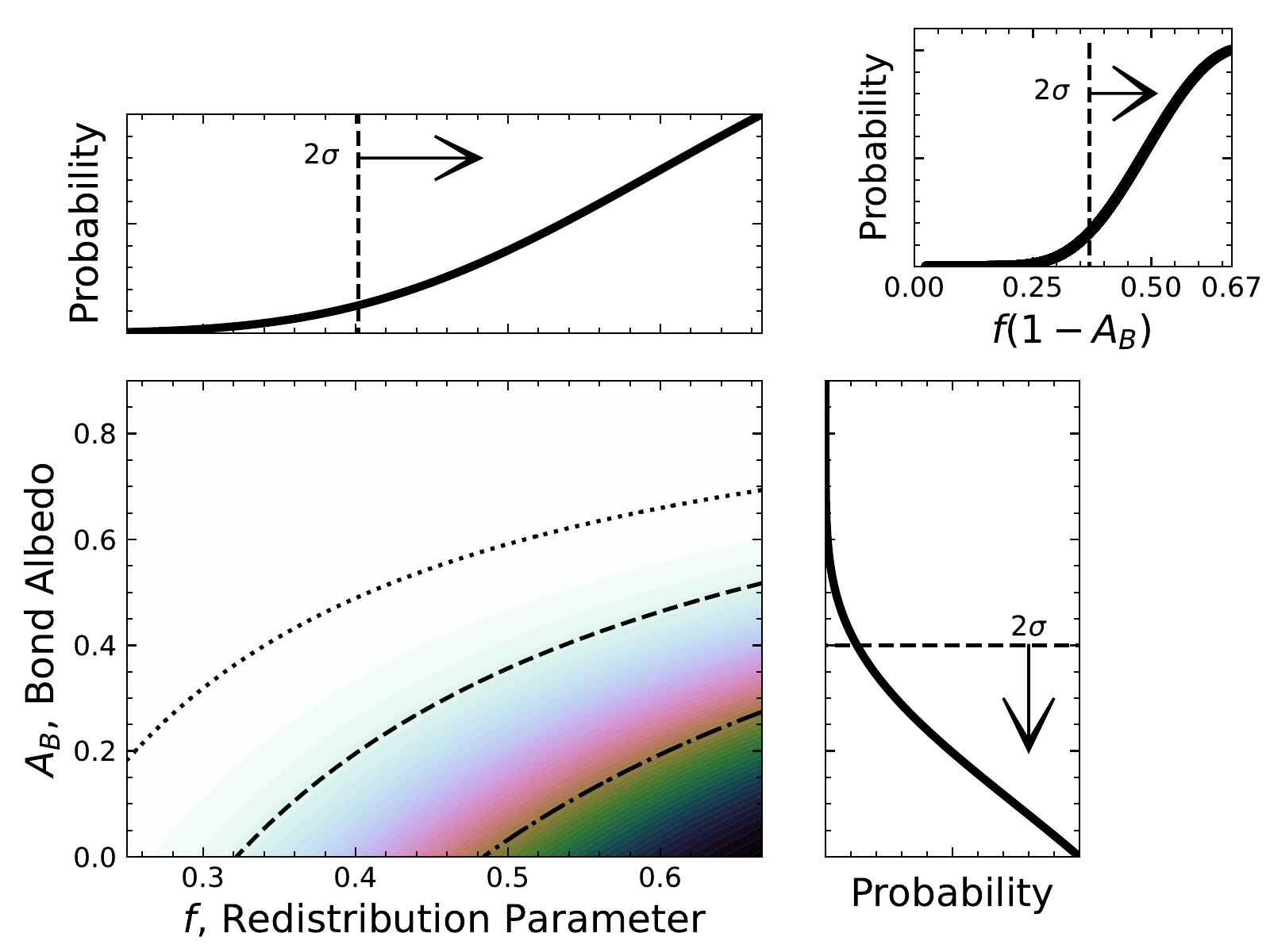}
\caption{\label{fig:fab} Joint constraints on Bond albedo and the heat
  redistribution parameter, assuming that \gjb\ radiates as a
  blackbody at its 4.5\,\micron\ brightness temperature. More heavily
  shaded regions indicate higher probabilities in the main panel, with
  the contour lines demarcating the enclosed 1$\sigma$, 2$\sigma$, and
  $3\sigma$ confidence intervals. The dashed lines in the 1D posterior
  distribution plots show the 95.4\% (2$\sigma$) confidence limit for
  the indicated parameter. Our measurements of \gjb\ constrain
  $f>$\,\fLL, $A_B<$\ \abUL, and $f(1-A_B)>$\ \fabLL. }
\end{center}
\end{figure}

One phenomenological framework for interpreting single-band secondary
eclipse measurements is to assume the planet radiates as a blackbody
at the measured brightness temperature, then to use the measurement to
constrain some combination of day-to-night heat redistribution
parameter $f$ and Bond Albedo $A_B$ \citep[e.g.,][]{seager:2010}. In
particular, the combination $f (1 - A_B)$ directly determines the
planet's day-side equilibrium temperature via
\begin{equation}
  T_\mathrm{eq} = T_\mathrm{eff} \sqrt{\frac{R_*}{a}} \left( f \left[ 1 - A_B \right] \right)^{1/4} .
  \label{eq:teq}
\end{equation}
In this formulation, the limiting values of $f$ are $\frac{2}{3}$,
indicating no heat redistribution (e.g., consistent with no
atmosphere) and $\frac{1}{4}$, indicating uniform heat redistribution
around the planet.  Fig.~\ref{fig:fab} shows our joint constraints on
$f$ and $A_B$, assuming flat priors on both quantities and the system
parameters listed in Table~\ref{tab:params}. In all cases the most
likely values are the ones that give the highest day-side
$T_\mathrm{eq}$, i.e.\ $f=\frac{2}{3}$ and $A_B = 0$. We set upper
limits (at 95.4\%, or 2$\sigma$ confidence) of: a low albedo of
$A_B<$\ \abUL, a high redistribution parameter of $f>$\,\fLL , and
a high combination of the two, $f(1-A_B)>$\ \fabLL. 

\subsubsection{One-dimensional Atmosphere Models}
\label{sec:radtrans}

Finally, we also present a large suite of atmospheric models and
spectra of \gjb.  These models are all available as machine-readable
supplements to this paper.  {\update Our models and spectra are generated using the
open-source, 1D radiative transfer code \texttt{HELIOS}
\citep{malik:2017, malik:2019a, malik:2019b}, which simulates the
planet in radiative-convective equilibrium and also provides
the functionality to include the radiative effects of a non-gray
surface} on both the atmosphere and the planetary spectrum (see
Whittaker et al., in prep.).

{\update In \texttt{HELIOS} the temperature profile and surface
  temperature are obtained using the k-distribution method, with 420
  wavelength bins (0.245 --- 10$^5$ $\mu$m). Then, starting from the
  equilibrium temperature profile, the planetary spectrum is
  calculated using opacity sampling with a resolution of $R$ =
  4000. Convectively unstable atmospheric layers are corrected using
  convective adjustment. We model dayside-averaged conditions and use
  the scaling theory of \citet{koll:2022} to estimate the amount of
  heat transported from the day-side to the night-side of the
  planet. In the bare-rock case, the heat redistribution parameter
  ($f$ in Eq.~\ref{eq:teq}) is set to $2/3$, equivalent to no
  horizontal heat transport \citep{burrows:2008, hansen:2008}. }

Gaseous opacities are calculated with \texttt{HELIOS-K}
\citep{grimm:2015, grimm:2021}, including O$_2$ \citep{gordon:2017},
N$_2$\citep{gordon:2022}, H$_2$O \citep{polyansky:2018}, CO
\citep{li:2015}, CO$_2$ \citep{rothman:2010}, CH$_4$
\citep{yurchenko:2017} and SO$_2$ \citep{underwood:2016}. All
opacities are calculated on a fixed grid with a resolution of 0.01
cm$^{-1}$, assuming a Voigt profile truncated at 100~cm$^{-1}$ from
line center. For H$_2$O, CH$_4$ and SO$_2$ the default pressure
broadening coefficients provided by the Exomol
database\footnote{https://exomol.com/data/molecules/} are
included. For O$_2$, N$_2$, CO and CO$_2$ the HITRAN broadening
formalism for self-broadening is used. Further included are collision
induced absorption (CIA) by O$_2$-O$_2$, O$_2$-CO$_2$, CO$_2$-CO$_2$,
N$_2$-N$_2$, and N$_2$-CH$_4$ pairs \citep{richard:2012} and Rayleigh
scattering of H$_2$O, O$_2$, N$_2$, CO$_2$ and CO \citep{cox:2000,
  sneep:2005, wagner:2008, thalman:2014}.

{\update To model the radiative effects of the surface, we use the
  geometric albedo spectra from \citet{hu:2012}.  For the bare-rock
scenario, \texttt{HELIOS} again} iterates the surface temperature until
the surface is in radiative equilibrium, i.e., the downward stellar
radiation equals the reflected plus emitted radiation at the surface
boundary. This takes the non-gray surface albedo into account across
the range of 0.3--25\,\micron, thus correctly treating both the
stellar flux absorption and reflection as well as the planetary
emission.

When modeling the planetary envelope we include the main infrared
absorbers that may be plausibly found in secondary atmospheres ---
\water, \coo, CO, \methane, and SO$_2$ \citep[see,
  e.g.,][]{gillmann:2022} --- and vary their mixing ratios between
1~ppm and 1\% \citep{gaillard:2014}. As background gas we use O$_2$ or
N$_2$ \citep{wordsworth:2014,luger:2015,schaefer:2016,lammer:2019}. As
limiting cases we also approximate post-water-runaway, Venus-like and
carbon-rich (elemental C/O $\sim$ 1) atmospheres by adding pure
\water, \coo, and CO scenarios
\citep{goldblatt:2015,kane:2014,madhusudhan:2012b}.  Specifically, the
full range of our models included either CO or CH$_4$ in an
N$_2$-dominated atmosphere; CO$_2$, SO$_2$, or H$_2$O in an
O$_2$-dominated atmosphere; and atmospheres of pure CO$_2$, SO$_2$, or
H$_2$O. These models are not intended to be exhaustive (nor may they
all be chemically stable on geological timescales); our goal is to
explore a representative range of atmospheres without overinterpreting
our single-channel measurement.

\begin{figure}[bt!]
\begin{center}
\includegraphics[width=3.5in]{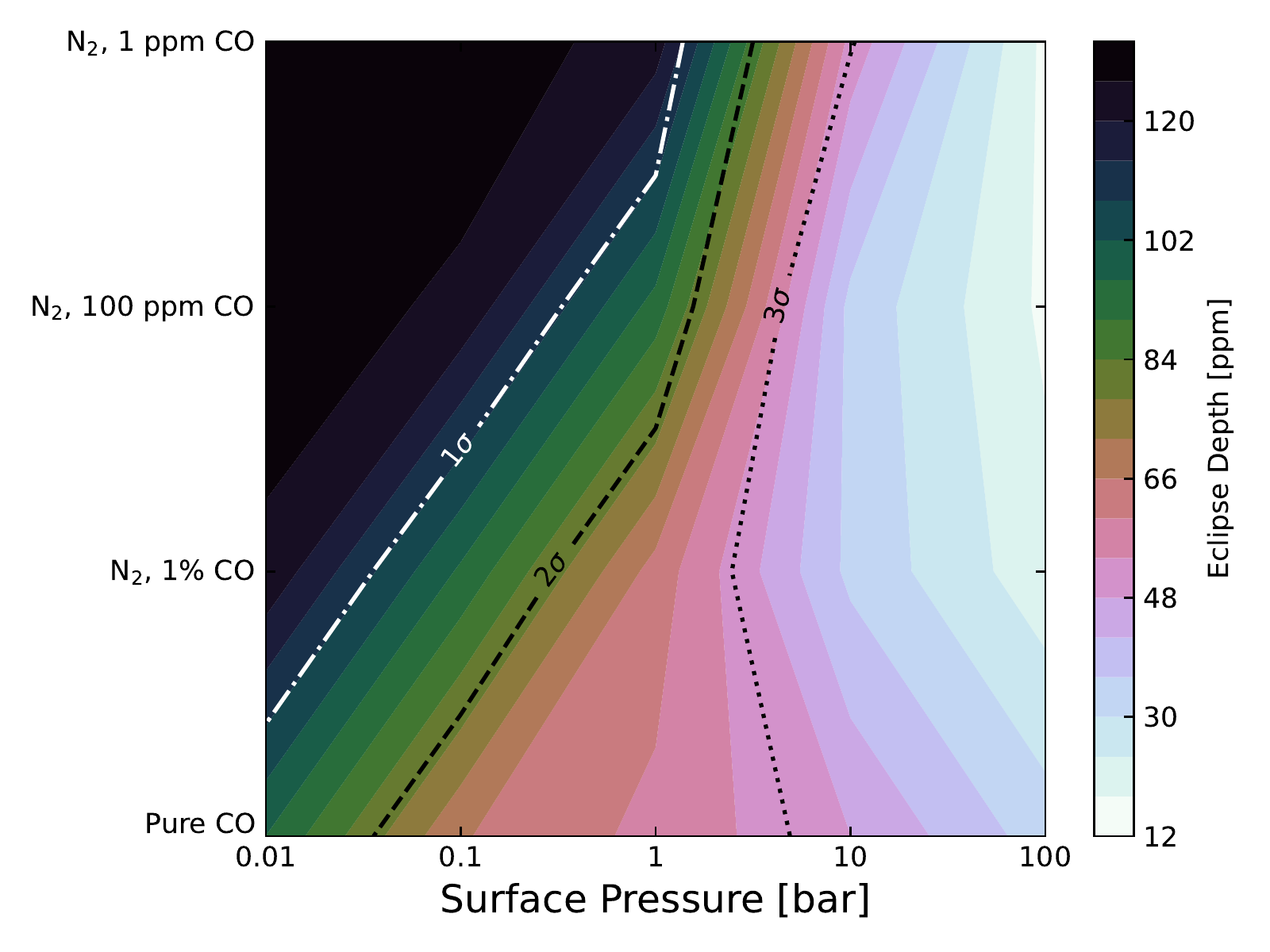}
\includegraphics[width=3.5in]{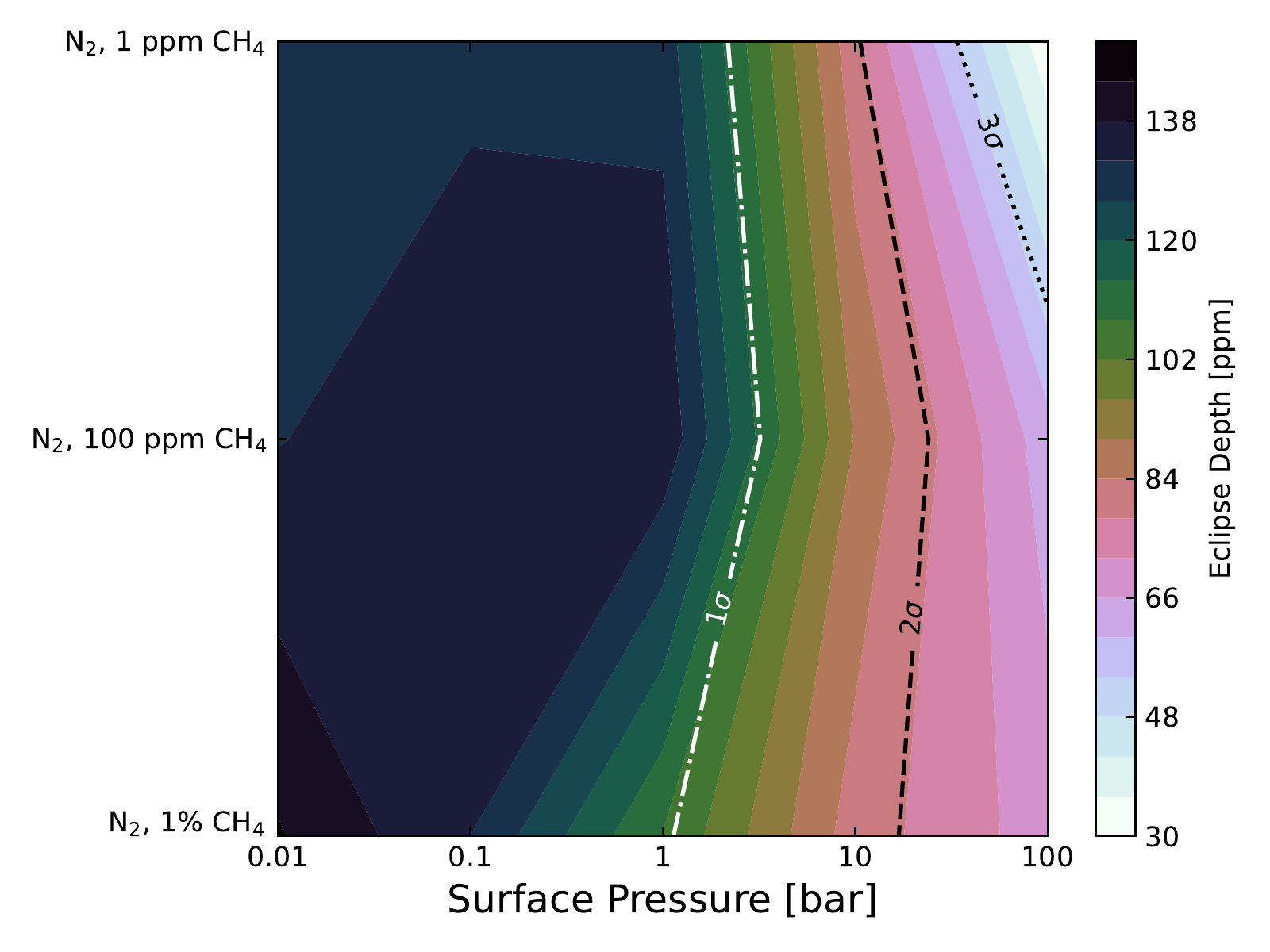}
\caption{\label{fig:n2_contours} Predicted IRAC2 4.5\,\micron\ eclipse
  depths (colored shading) from our suite of atmospheric models for
  atmospheres dominated by N$_2$; each model has a single dominant,
  active IR absorber of either CO ({\em top}) or CH$_4$ ({\em
    bottom}). The contour lines indicate the level of
  agreement with our eclipse measurement of \depth\udepth, with darker regions being more
  consistent with our data.  Thick atmospheres with $P_\mathrm{surf} \gtrsim 10$~bar are disfavored by all modeled  mixing ratios. }
\end{center}
\end{figure}

\begin{figure}[bt!]
\begin{center}
\includegraphics[width=3.55in]{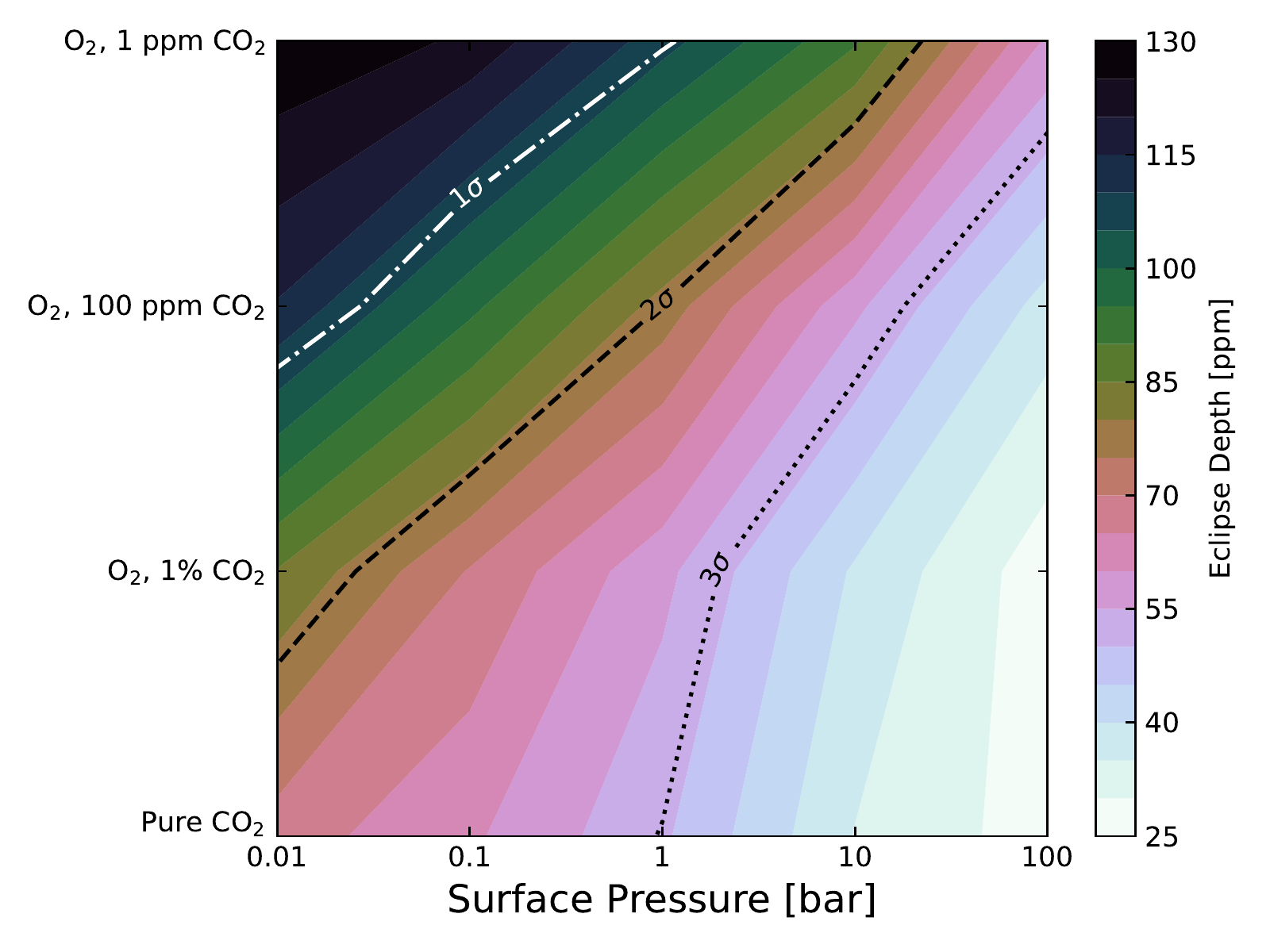}
\includegraphics[width=3.55in]{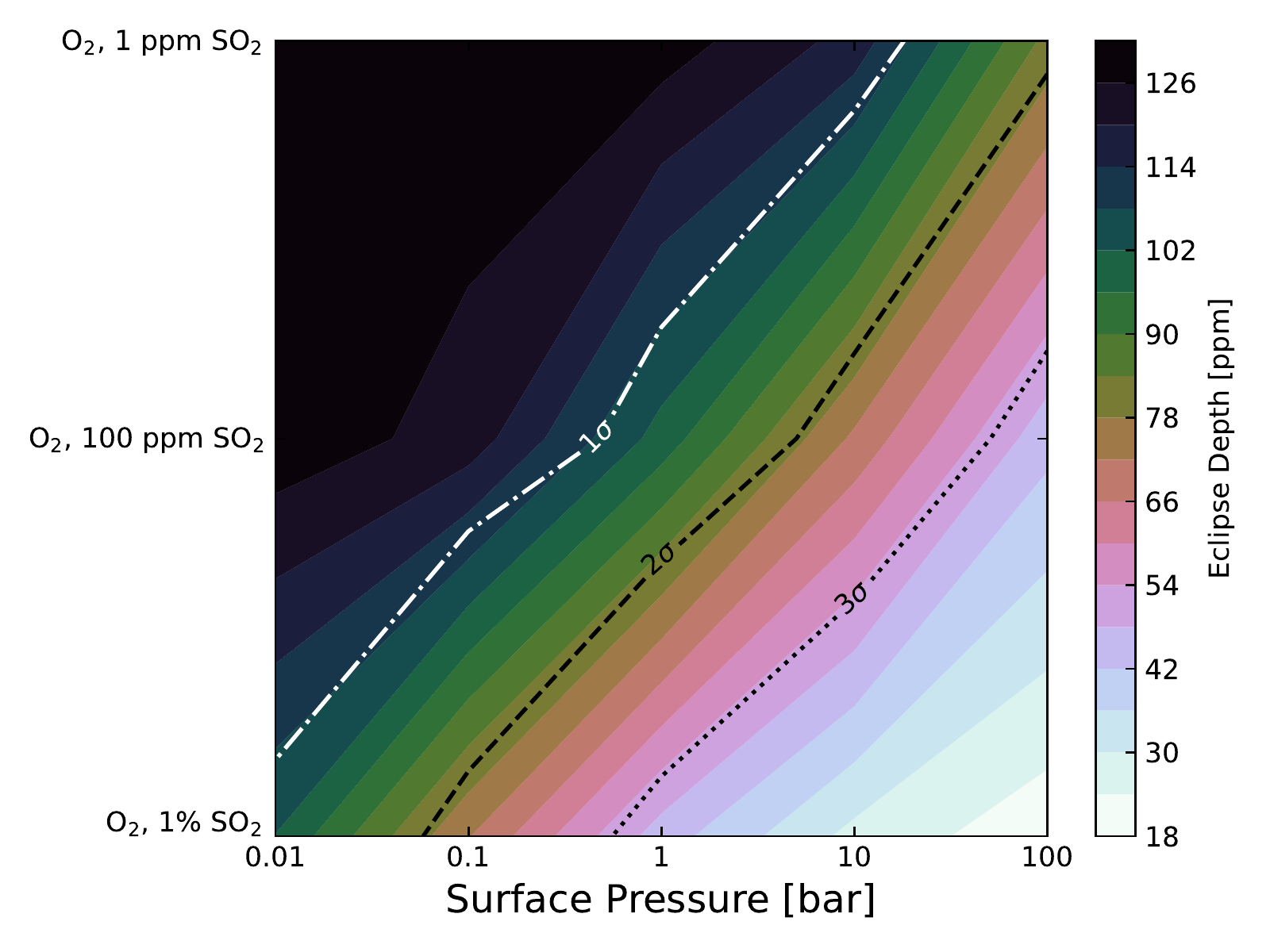}
\includegraphics[width=3.55in]{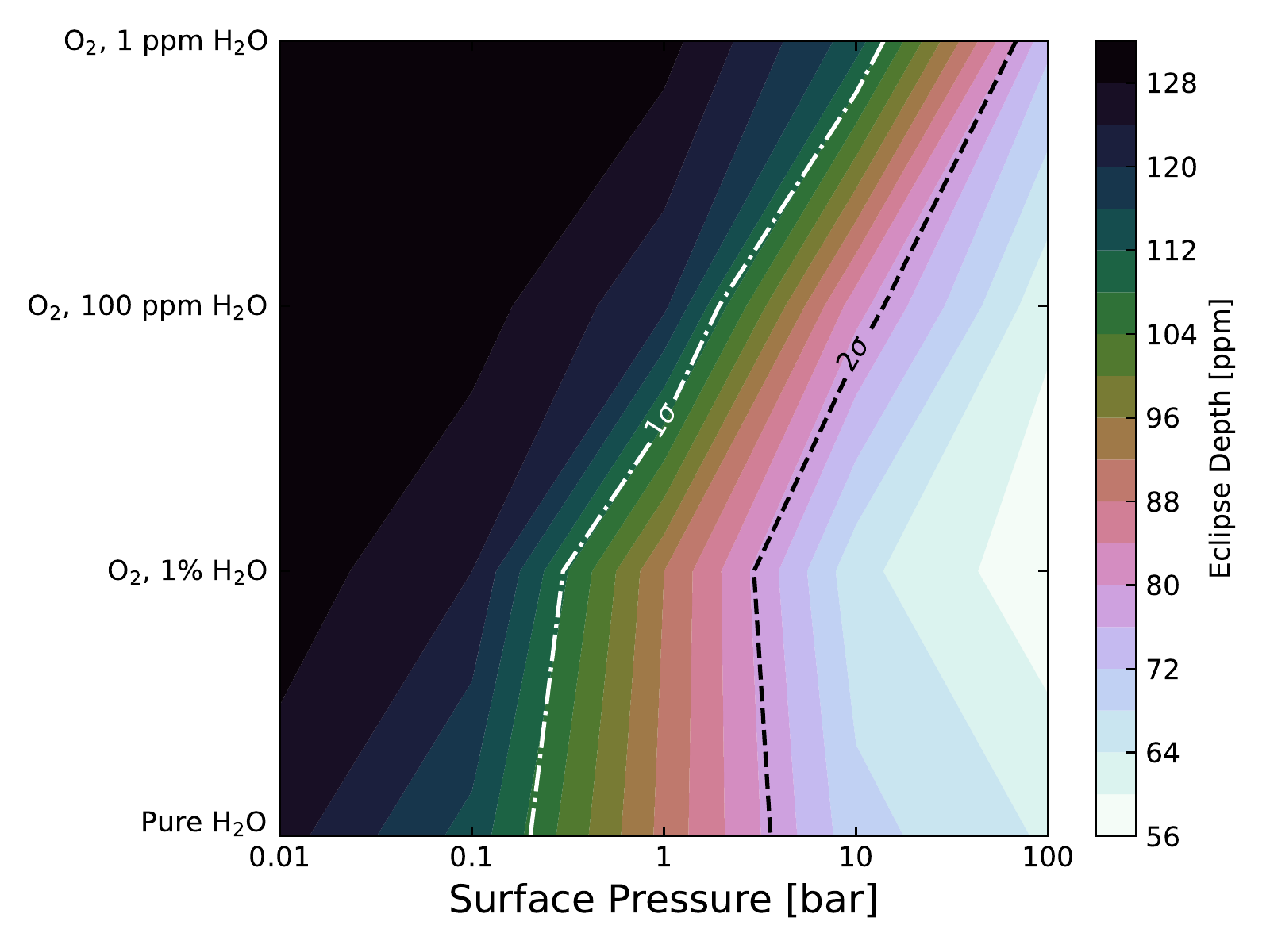}
\caption{\label{fig:o2_contours} Same as Fig.~\ref{fig:n2_contours}
  but for atmospheres dominated by O$_2$, and with active IR absorbers
  of \coo\ ({\em top}), SO$_2$ ({\em middle}), and \water\ ({\em
    bottom}). Thick atmospheres ($P_\mathrm{surf} \gtrsim 10$~bar) with mixing ratios  of
  $\gtrsim 10^{-4}$ are disfavored in all cases. }
\end{center}
\end{figure}

For each model emission spectrum, we calculated the eclipse depth that
would be measured in the 4.5\,\micron\ IRAC2 bandpass.
Figs.~\ref{fig:n2_contours} and~\ref{fig:o2_contours} compare our
measured 4.5\,\micron\ eclipse depths to the predictions of our
atmospheric models dominated by N$_2$ and O$_2$, respectively.  For
the N$_2$-dominated models, Fig.~\ref{fig:n2_contours} shows that the
only models consistent with our eclipse measurement at 2$\sigma$ or
better have $P_\mathrm{surf}\lesssim 1$~bar (for CO as the active IR
absorber) and $\lesssim 10$~bar (for \methane).  Similarly,
Fig.~\ref{fig:o2_contours} shows that the O$_2$-dominated models
consistent with our measurement at 2$\sigma$ have $P_\mathrm{surf}
\lesssim 1$~bar (for \coo) and $\lesssim 10$~bar (for SO$_2$ or
\water).

\begin{figure}[bt!]
\begin{center}
  \includegraphics[width=3.4in]{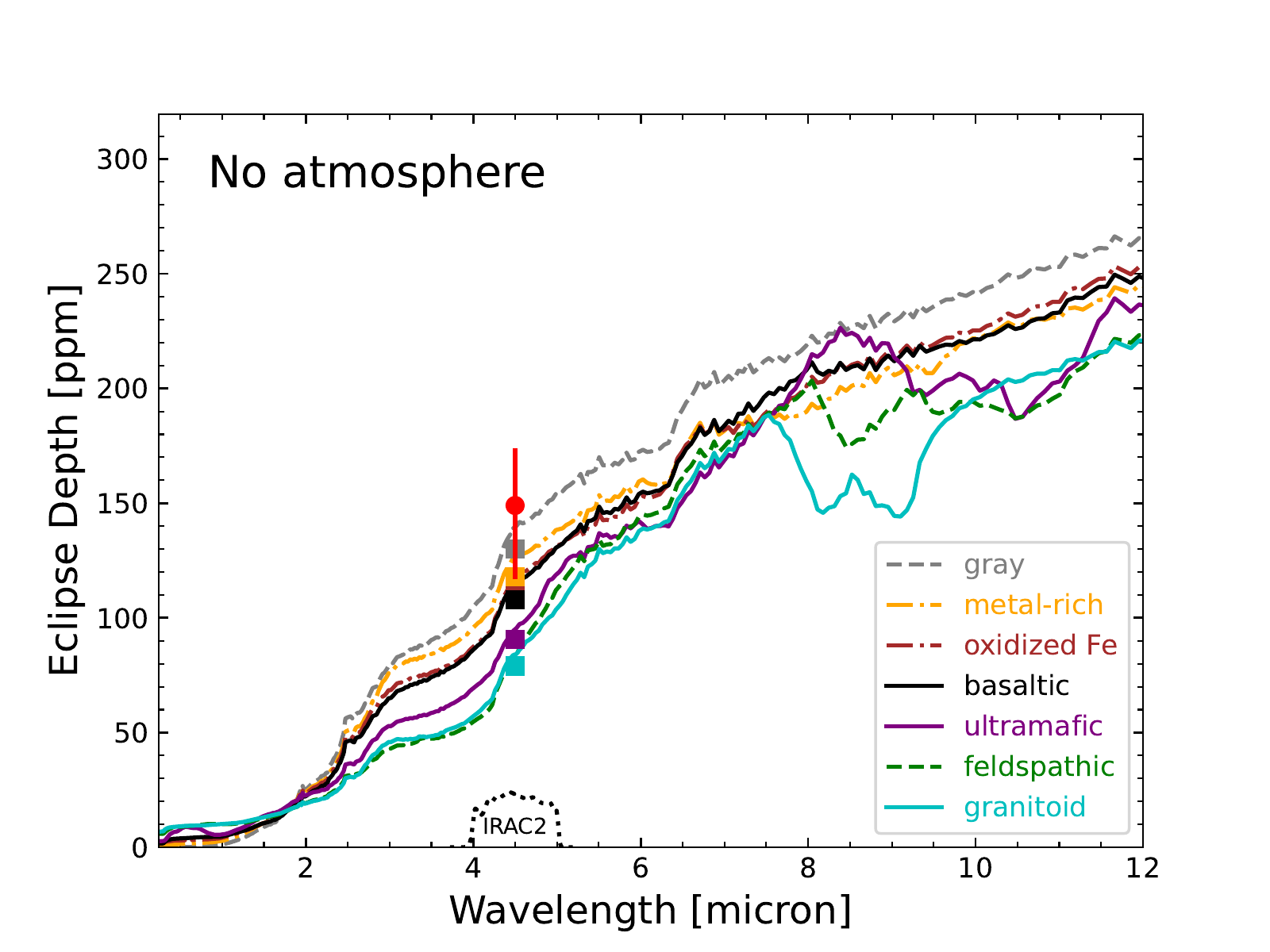}
\caption{\label{fig:rockspec} Simulated eclipse spectra of \gjb\ from
  our suite of no-atmosphere models for different types of solid,
  unmelted surface mineralogies \citep[lines; from][]{hu:2012}
  compared to our measured depth and uncertainty (red circle with
  error bars).  Squares show the values obtained by averaging the
  models over the IRAC2 bandpass (dotted line).  All but the granitoid
  and feldspathic models are consistent with our measurement at
  $\le$2$\sigma$.}
\end{center}
\end{figure}

We also find that models lacking any atmosphere are also consistent
with our eclipse measurement.  Fig.~\ref{fig:rockspec} shows that
while only the gray and metal-rich bare-rock models are consistent
with our measurement at $<$1$\sigma$, models with an oxidized Fe,
basaltic, and ultramafic surface composition are all consistent with
our eclipse at $<$2$\sigma$.   Models with a feldspathic or
granitoid surface composition are inconsistent with the data at
$>$2$\sigma$. However, we note that  the substellar region
of \gjb\ is likely hot enough for such materials to melt. Although our
atmosphere-free models may not be strictly accurate for this planet's
surface, we leave more detailed models involving both solid and melted
regions for  future study.

\section{Atmospheric Evolution and Escape}
\label{sec:escape}
\cite{zahnle:2017} postulate a ``cosmic shoreline'' in which the
ability of a body to retain an atmosphere depends on some combination
of its escape velocity $v_\mathrm{esc}$, bolometric irradiation $S$,
and extreme UV (XUV) irradiation.  Given its irradiation and
$v_\mathrm{esc}$=\vesc\uvesc~m~s$^{-1}$, \gjb\ would lie well into the
atmosphere-free zone --- but then so would 55~Cnc~e, where infrared
measurements seem to indicate the presence of an atmosphere
\citep{demory:2016,tamburo:2018,demory:2016b,angelo:2017,hammond:2017}. Thus the cosmic shoreline may not
apply universally to smaller exoplanets irradiated so much more
intensely than anything in the Solar System.

The evolution of a planetary atmosphere into its end-state depends on
the mantle's cooling rate, regions of oxidization, and potential sinks
for these volatiles \citep[e.g.,][]{gillmann:2022}. For example, N$_2$
is interesting because there are relatively few sinks and so, despite
N$_2$ being a relatively small amount of the Earth's total volatile
inventory, it dominates the present atmosphere. Another interesting
species to think about is SO$_2$: the sulfur cycle on Venus is an
important component of its overall atmospheric chemistry, and so
SO$_2$ is a significant component of volcanic outgassing on Venus
\citep{esposito:1984,korenaga:2010,zhang:2012,kane:2019}. Nonetheless SO$_2$ does
not constitute nearly as much of the Venusian atmosphere as one might
expect, largely because it reacts with calcium carbonates to produce
CO \citep{hong:1997}. Thus one possible end-state for a desiccated
rocky planet's atmosphere would be an atmosphere consisting of mainly
\coo, N$_2$, CO, CH$_4$, and SO$_2$. Any \water\ would remain through
the moist greenhouse phase (if any), but would probably end up the
same fate as past water on Venus: disassociation, loss of H$_2$, and
oxidization of the surface and reaction with CH$_4$ to produce more
\coo\ \citep{kane:2020}.

\subsection{Energy-Limited Escape}
We first estimate the atmospheric loss rate from \gjb\ using the
formalism of energy-limited atmospheric escape \citep{salz:2016},
leaving more involved estimates of the planet's mass-loss rate for
future work.  Using the MUSCLES treasury
survey's spectra of nearby M dwarfs
\citep{france:2016,youngblood:2016,loyd:2016} we estimate an XUV flux
incident on \gjb\ of $(6-8) \times
10^3$~erg~s$^{-1}$~cm$^{-2}$. Assuming a heating efficiency of 0.3
\citep{salz:2015} and that the planet's optical transit radius is the
same as its effective radius of XUV absorption, this XUV flux
translates into an atmospheric mass loss rate of
roughly $0.1 M_\oplus$~Gyr$^{-1}$.  The mass of a planet's atmosphere
is just $4\pi R_p^2 P_\mathrm{surf} / g_\mathrm{surf}$, which for
\gjb\ is $\left( 5.3 \times 10^{18}\mathrm{~kg} \right)\left(
P_\mathrm{surf}/\mathrm{1~bar} \right)$.  Thus even a 100~bar
atmosphere would be ablated in $<1$~Myr. Note that although
\gj\ exhibited no detectable stellar flares during its first sector of
\tess\ observations \citep{howard:2022} the star was presumably more
active, and thus mass-loss rates from \gjb\ would have been even
higher, earlier in the system's lifetime.

\subsection{Comparing Outgassing and Escape Rates}

\begin{figure}[bt!]
\begin{center}
\includegraphics[width=3.2in]{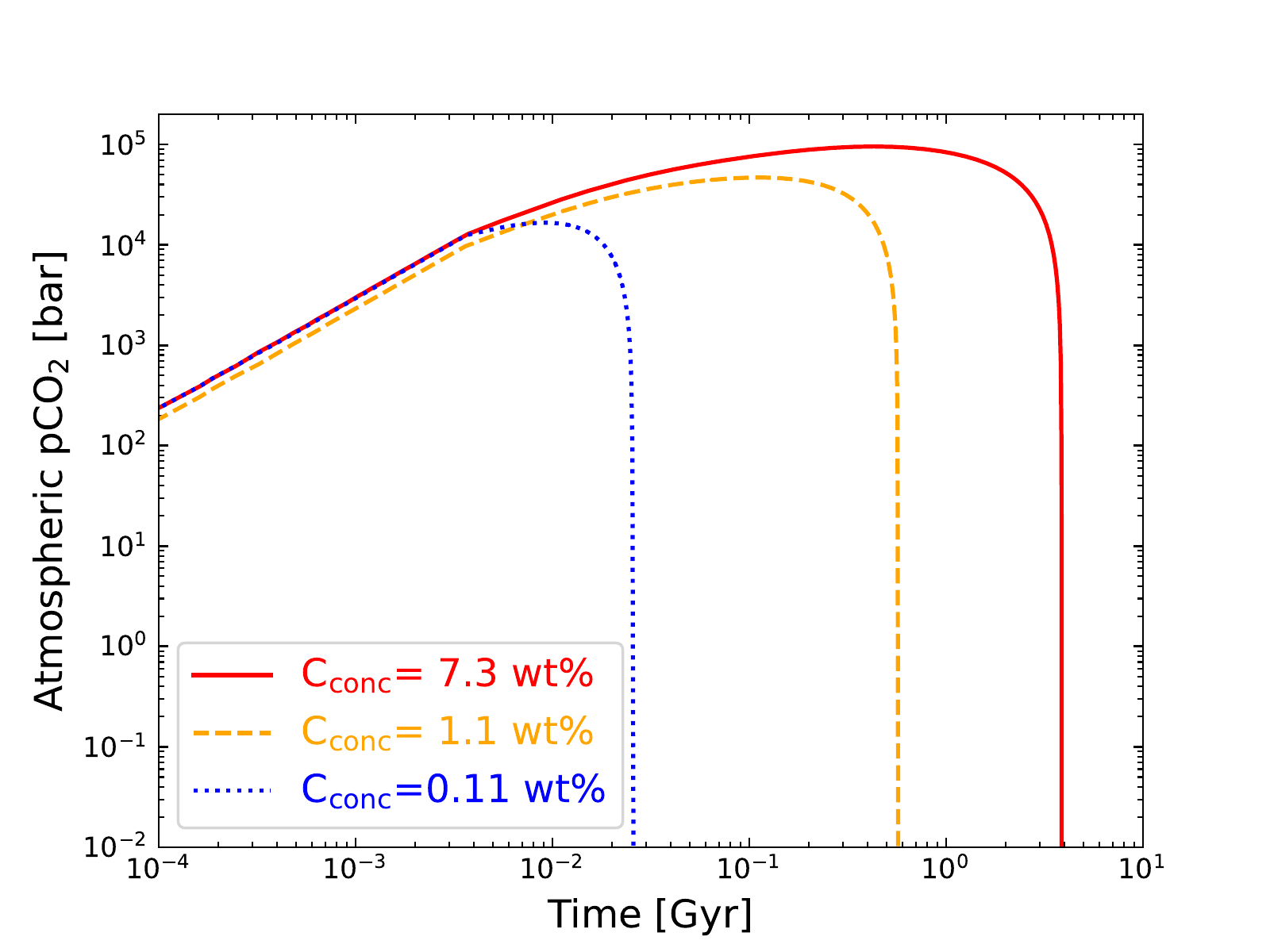}
\caption{\label{fig:carbon} Time evolution of  atmospheric
  pressure with variations in mantle carbon concentration
  (C$_\mathrm{conc}$) for models assuming an otherwise Earth-like
  core. An Earth-like initial carbon concentration of 0.011 wt\% (not
  shown) results in an atmospheric mass loss rate always greater than
  the outgassing rate, so no atmosphere can build up.  Models with
  greater carbon inventories than found on Earth build up temporary
  secondary atmospheres that are nonetheless lost within 4~Gyr. Even
  with orders of magnitude greater initial carbon inventory than
  Earth, GJ~1252b would have no significant remaining atmosphere
  due to its prodigious atmospheric loss rate.  }
\end{center}
\end{figure}

To further evaluate the prospects for volatile loss from \gjb\, we
modify the model of \cite{foley:2018,foley:2019} to apply to GJ~1252b and explore
varying initial mantle \coo\ inventories that will still allow for a
completely desiccated planet at the estimated planetary age of
$\sim$3.9~Gyr. Mantle gravity and core-radius-fraction model inputs
were calculated using the mass-radius-composition solver,
\texttt{ExoPlex} \citep{unterborn:2019}. As the planet lacks any
significant volatile atmosphere, we assumed the planet was made
entirely of a FeO-free silicate mantle and pure-liquid-Fe core. Using
a planet radius of 1.213\,$R_\oplus$, mass of {\update 1.32\,$M_\oplus$} and
assuming an Earth-like core radius fraction of 0.55\%,
\texttt{ExoPlex} calculates an average mantle density of
5026~kg~m$^{-3}$ and gravity 14.4~m~s$^{-2}$. Due to the high surface
temperature of the planet and the low likelihood of liquid water on
the surface, we assume that the planet is in the stagnant lid regime
of tectonics \citep[e.g., as is Venus;][]{gillmann:2022}. Our model then
assumes all \coo\ outgassed from the mantle will accumulate in the
atmosphere, as there is no known method of weathering or recycling
carbon without water \citep{walker:1981,  kasting:2003,
  foley:2016}. Our model also assumes an Earth-like heat producing
element budget, initial mantle temperature of 2000~K, and reference
viscosity $\mu_\mathrm{ref} = 1.3 \times 10^{20} \mathrm{~Pa~s}$
\citep{foley:2018}.

Assuming an atmosphere mass loss rate of $5 \times
10^6$ kg~s$^{-1}$, we vary the initial mantle \coo\ inventory of our
model between an Earth-like initial inventory of 0.011~wt\% (mass
percentage), based on an estimate of $10^{22}$ mol of \coo\ in the
mantle and surface reservoirs of Earth \citep{sleep:2001}, to 7.3
wt\%\ (over two orders of magnitude greater \coo\ than on present-day
Earth, by mass fraction).

Fig.~\ref{fig:carbon} shows the time evolution of total atmospheric
pressure with varying initial mantle \coo\ inventories. For the larger
carbon inventories considered, the planet's initial rapid outgassing
is greater than the atmospheric loss rate, allowing the atmospheric
pressure to build before gradually being eroded once the mantle's
carbon store is depleted. However, for the Earth-like initial
\coo\ ($1.1 \times 10^{-2}$ wt\%) the atmospheric loss rate is always
greater than the outgassing rate and so the planet never builds a
significant secondary atmosphere. All models tested result in a planet
with a completely eroded atmosphere by 3.9~Gyr. Thus GJ~1252b may have
had orders of magnitude greater carbon inventory than Earth yet still
have no remaining atmosphere today.

Taken together, our model spectra and escape calculations strongly
indicate that \gjb\ has no significant atmosphere.

\section{Discussion and Conclusions }
\label{sec:conclusions}

\begin{figure*}[bt!]
\begin{center}
\includegraphics[width=3.2in]{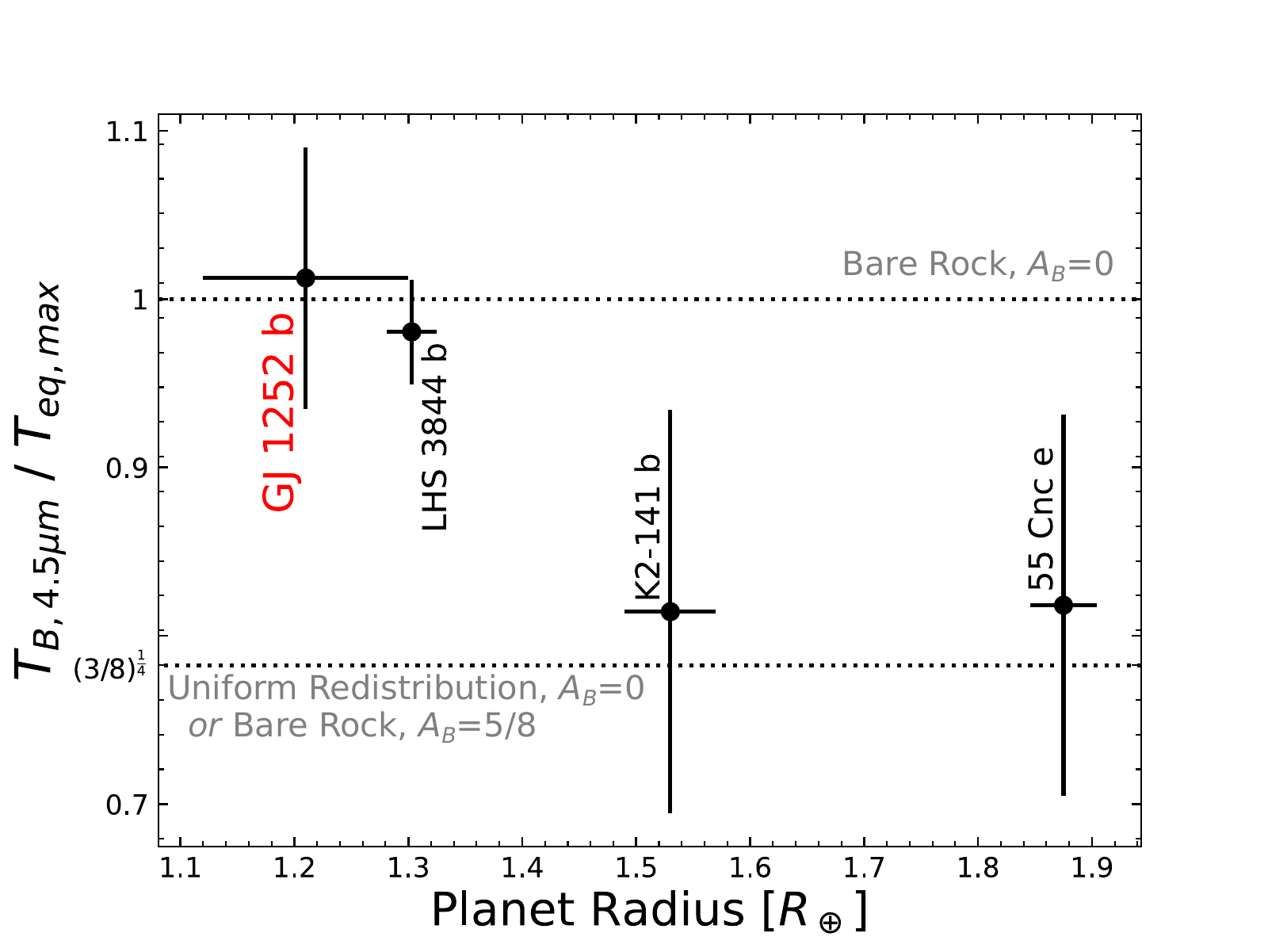}
\includegraphics[width=3.2in]{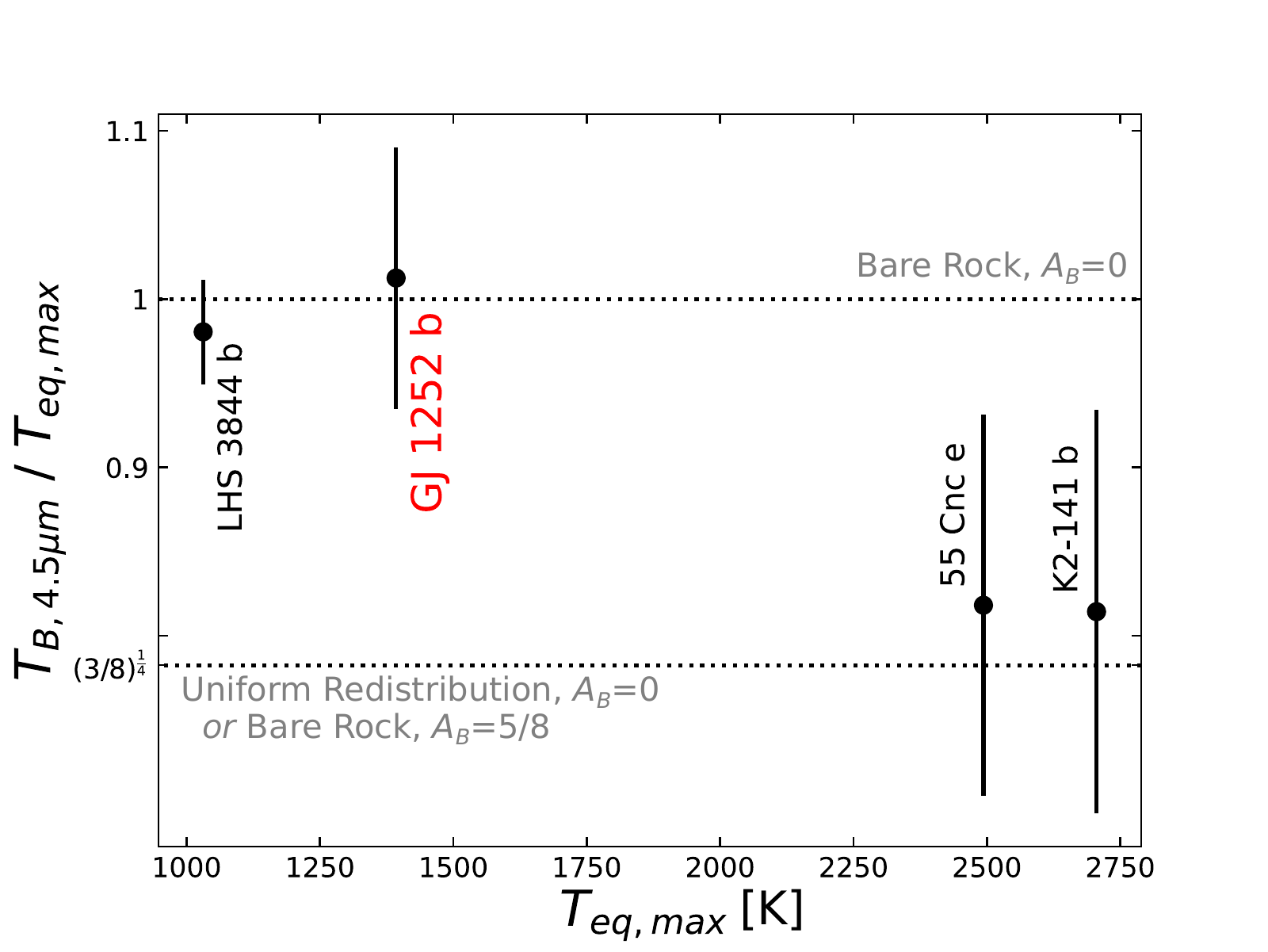}
\caption{\label{fig:comparison}
Normalized day-side brightness temperatures, $T_{B,4.5\,\micron}$, of
all super-earths with infrared flux measurements plotted vs.\ planet
radius ({\em left}) and $T_\mathrm{eq,max}$ ({\em right}).  The two smaller,
cooler planets have 4.5\,\micron\ brightness temperatures consistent
with no atmosphere and zero albedo while the larger, hotter planets
are significantly cooler and consistent with uniform redistribution of
incident radiation and/or nonzero albedo. }
\end{center}
\end{figure*}

\subsection{Comparisons With Similar Exoplanets}
\gjb\ joins the handful of small planets ($<2 R_\oplus$) with infrared
flux detections. The prior examples are 55~Cnc~e (whose size and mass
imply a sizable volatile mass fraction), LHS~3844b, and
K2-141b\footnote{A similar measurement was also recently reported for
  TOI-824b (Roy et al., in press), but that planet's bulk density
  clearly classifies it as a volatile-rich sub-Neptune rather than a
  rocky planet \citep{burt:2020}.}; some relevant parameters for all
these systems (including \gjb) are listed in Table~\ref{tab:hotrocks}.
\gjb\ is smaller than all of these other planets but intermediate in
irradiation.  We used the system parameters in
Table~\ref{tab:hotrocks}, along with BT-Settl model stellar spectra
\citep{allard:2014} interpolated to these stars' parameters, to
homogeneously calculate the irradiation and 4.5\,\micron\ day-side
brightness temperatures of all these planets. The model spectra used
for all four planets are included as machine-readable files with this paper.

As shown in Fig.~\ref{fig:comparison}, a curious dichotomy
emerges. The two smallest and coolest planets (\gjb\ and LHS~3844b)
both have day-side brightness temperatures $\left( T_{B,4.5\micron}
\right)$ consistent with the maximum possible day-size equilibrium
temperature (i.e., $f=\frac{2}{3}$ and $A_B=0$ in Eq.~\ref{eq:teq}):
\begin{equation}
  T_\mathrm{eq,max} = \left( \frac{2}{3} \right)^{1/4} T_\mathrm{eff} \sqrt{\frac{R_*}{a}} .
\end{equation}
On the other hand, the two largest and hottest planets (55~Cnc~e and
K2-141b) both have notably lower normalized day-side temperatures that
are consistent with uniform heat redistribution ($f=\frac{1}{4}$).

This dichotomy may be coincidental: 55~Cnc~e's emission measurements
are best interpreted as indicating a massive atmosphere, while the
interpretation for K2-141b is a nonzero albedo but negligible
atmosphere \citep{zieba:2022}.  Some calculations predict that rocky
planets with sufficiently intense irradiation could exhibit a high
substellar albedo induced by photovolatilization of day-side rocky
materials \citep{kite:2016,mansfield:2019}. One possibility is
therefore that there is an irradiation threshold for this albedo
enhancement, with \gjb\ and LHS~3844b lying below it and K2-141b lying
above it.

Alternatively, both 55~Cnc~e and K2-141b may have atmospheres thick
enough that they can transport sufficient heat to measurably cool
their day-sides. Evidence of 55~Cnc~e's thick atmosphere is seen in
its asymmetric 4.5\,\micron\ phase curve \citep{demory:2016b}, but
K2-141b's 4.5\,\micron\ phase curve showed no such evidence for a
thick atmosphere \citep{zieba:2022} -- this despite K2-141b's day-side
being heated high enough above the silicate solidus that an optically
thick, $\sim$0.1~bar mineral atmosphere is predicted. Future modeling
and observations will both be needed to determine whether
Fig.~\ref{fig:comparison} represents a coherent trend between some
combination of irradiation, planet size, heat redistribution, and
albedo.

Fortunately, \gj\ is bright enough to offer high S/N while faint
enough to be observable with all of {\em JWST}'s instruments.  With an
Emission Spectroscopy Metric \citep[ESM;][]{kempton:2018} greater than
K2-141b and within a factor of two of LHS~3844b (see
Table~\ref{tab:hotrocks}), \gjb\ is likely to join the select group
of eminently observable and highly irradiated terrestrial exoplanets
and to be subjected to many productive future investigations.

\begin{deluxetable}{l l l l  l}[bt!]
\tabletypesize{\scriptsize}
\tablecaption{\small Planets $<2 R_\oplus$ and With Measured Infrared Emission   \label{tab:hotrocks}}
\tablewidth{0pt}
\tablehead{
\colhead{Parameter} & \colhead{GJ~1252b\tablenotemark{a}} & \colhead{LHS 3844b\tablenotemark{b}} & \colhead{K2-141 b\tablenotemark{c}} & \colhead{55 Cnc e\tablenotemark{d}} 
}
\startdata
$R_p/R_\oplus$    & \rp & $1.32 $ & $1.53 $ & $1.88 $ \\
$a/R_*$          & $5.03$ & 7.08 & 2.36& 3.51 \\
$T_\mathrm{eff}$ / K  &  $3458$ & $3036$ & 4599 & $5172 $ \\
$T_\mathrm{eq,max}$ / K&  1390 & 1030 & 2705 & 2550\\
$\delta_{4.5\,\micron}$ / ppm & \depth\udepth  & $380 \pm 40$ & $143 \pm 39$ & $83 \pm 23$\tablenotemark{e}  \\
$T_{B, 4.5\,\micron}$ / K & \tday\utday & $1011^{+32}_{-33}$ & $2200^{+310}_{-330} $ & $2040^{+270}_{-290}$ \\
ESM$_{7.5\,\micron}$  & 16.7 & 30.0 & 15.4 & 70.1\tablenotemark{f}
\enddata
\tablenotetext{a}{From this work and \cite{shporer:2020}.}
\tablenotetext{b}{From \cite{vanderspek:2019} and \cite{kreidberg:2019}.}
\tablenotetext{c}{From \cite{malavolta:2018} and \cite{zieba:2022}.}
\tablenotetext{d}{From \cite{demory:2016} and \cite{bourrier:2018b}.}
\tablenotetext{e}{We adopt the standard deviation on the mean of the  eclipse depths of \cite{demory:2016}  as a conservative estimate of this uncertainty. {\update The weighted mean and its uncertainty from their first and second seasons of observations are $51\pm17$~ppm and $171\pm27$~ppm, respectively, corresponding to $1620 \pm 230$~K and $3070\pm300$.}}
\tablenotetext{f}{For stars as bright as 55~Cnc, ESM-like metrics typically overestimate the achievable S/N {\update \citep{kempton:2018}}.}
\end{deluxetable}

\subsection{Conclusions}

We have presented our measurement of 4.5\,\micron\ thermal emission
from the highly irradiated terrestrial planet \gjb. With a radius of
just \rp\,$R_\oplus$, our target is the smallest planet for which such
a measurement has been reported. After presenting our {\em Spitzer}
data analysis, along with an updated transit analysis using new
\tess\ data in Sec.~\ref{sec:obs}, Sec.~\ref{sec:interp} compared this
measurement to a large suite of atmospheric models and simulated
spectra.  Our modeling demonstrated that for a broad range of possible
atmospheric compositions surface pressures $P_\mathrm{surf}
\lesssim10$~bar are required to be consistent with the measured
eclipse at 2$\sigma$ confidence. Furthermore, Sec.~\ref{sec:escape}
then showed that the energy-limited atmospheric mass loss from
\gjb\ could  quickly erode atmospheres with
$P_\mathrm{surf}>100$~bar on timescales far shorter than the system
lifetime.

We therefore conclude that \gjb\ possesses no significant atmosphere.
In this case, it presumably retains only a tenuous mineral exosphere;
such an atmosphere would be expected to have $P_\mathrm{surf} \lesssim
10^{-6}$~bar and to be dominated by species such as Na, molecular
O$_2$ and atomic O, and K \citep{miguel:2011,ito:2015}.  Such an
atmosphere is likely thin enough that atmospheric circulation would
contribute negligibly to global heat and mass transfer
\citep[e.g.,][]{nguyen:2020}, although detailed simulations will be
needed to confirm this.  Since \gjb's atmosphere would be quite
optically thin, observations of \gjb\ therefore offer the opportunity
to directly probe surface minerology via emission spectroscopy during
eclipse and throughout the planet's orbit.

Infrared emission has been measured from only three exoplanets with
sizes placing them firmly in the terrestrial planet regime. A larger
sample is urgently needed to better identify the surface and
atmospheric properties of this class of planets; fortunately, this
number is already set to increase somewhat in the dawning JWST era
\citep[see discussion by][]{zieba:2022}.  A further pressing need is
to obtain more precise mass measurements for these planets: no mass
has been reported for LHS~3844b and \gjb\ has only a roughly 3$\sigma$
mass.  The combination of more precise masses and bulk densities, more
precise eclipse spectra, and stellar abundance measurements will
ultimately enable more accurate models to link these planets' surface
minerologies and atmospheres to their observed thermal emission.




\acknowledgments We thank the anonymous referee for their comments
that substantively improved the quality of this work. We also thank
Mike Werner, Katherine de Kleer, Daniel Koll, David Berardo, Courtney
Dressing, and Farisa Morales for productive discussions about this
project. We thank all the Spitzer Science Center staff for ensuring
that these observations, as well as the {\em Spitzer} mission, were
executed successfully. This work was supported in part by a grant from
NASA's Interdisciplinary Consortia for Astrobiology Research (ICAR).
MH would like to acknowledge NASA support via the FINESST Planetary
Science Division, NASA award number 80NSSC21K1536. Finally, IJMC
thanks JKC and NLC for providing considerable logistical support while
a large portion of the initial analysis was completed.

 \facility{TESS, Spitzer }

\bibliographystyle{apj}
\bibliography{../ms}

\appendix

\begin{figure*}[b]
\centering
\includegraphics[width=0.5\textwidth]{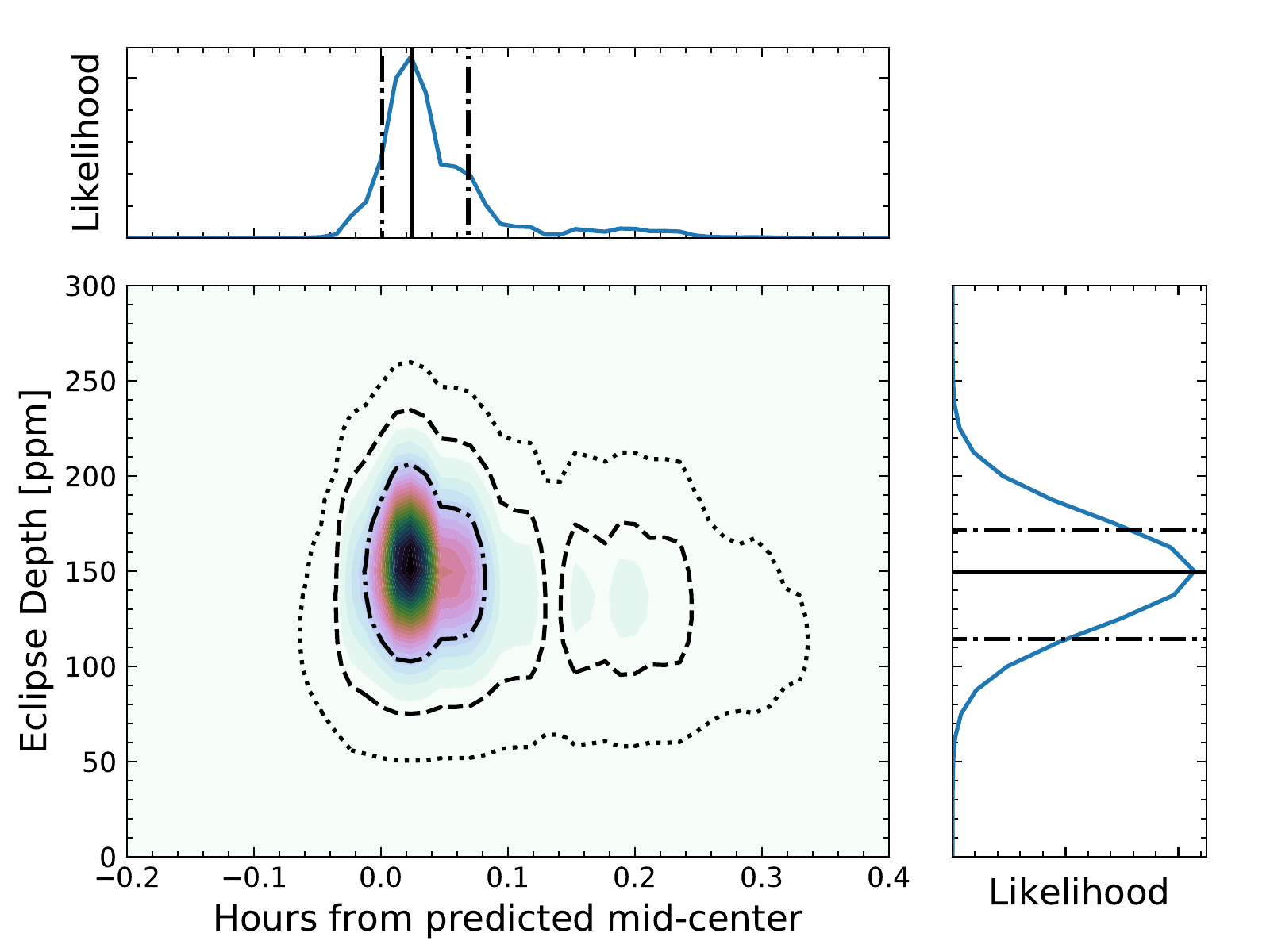}
\caption{\update Joint posterior distributions of eclipse depth and time of
  eclipse center from the analysis plotted in
  Fig.~\ref{fig:eclipse}. The eclipse depth exhibits no covariance
  with the  eclipse timing, despite the latter's long tail toward high values,
  \label{fig:corner}}
\end{figure*}

\begin{figure*}
\centering
\includegraphics[width=0.45\textwidth,page=1]{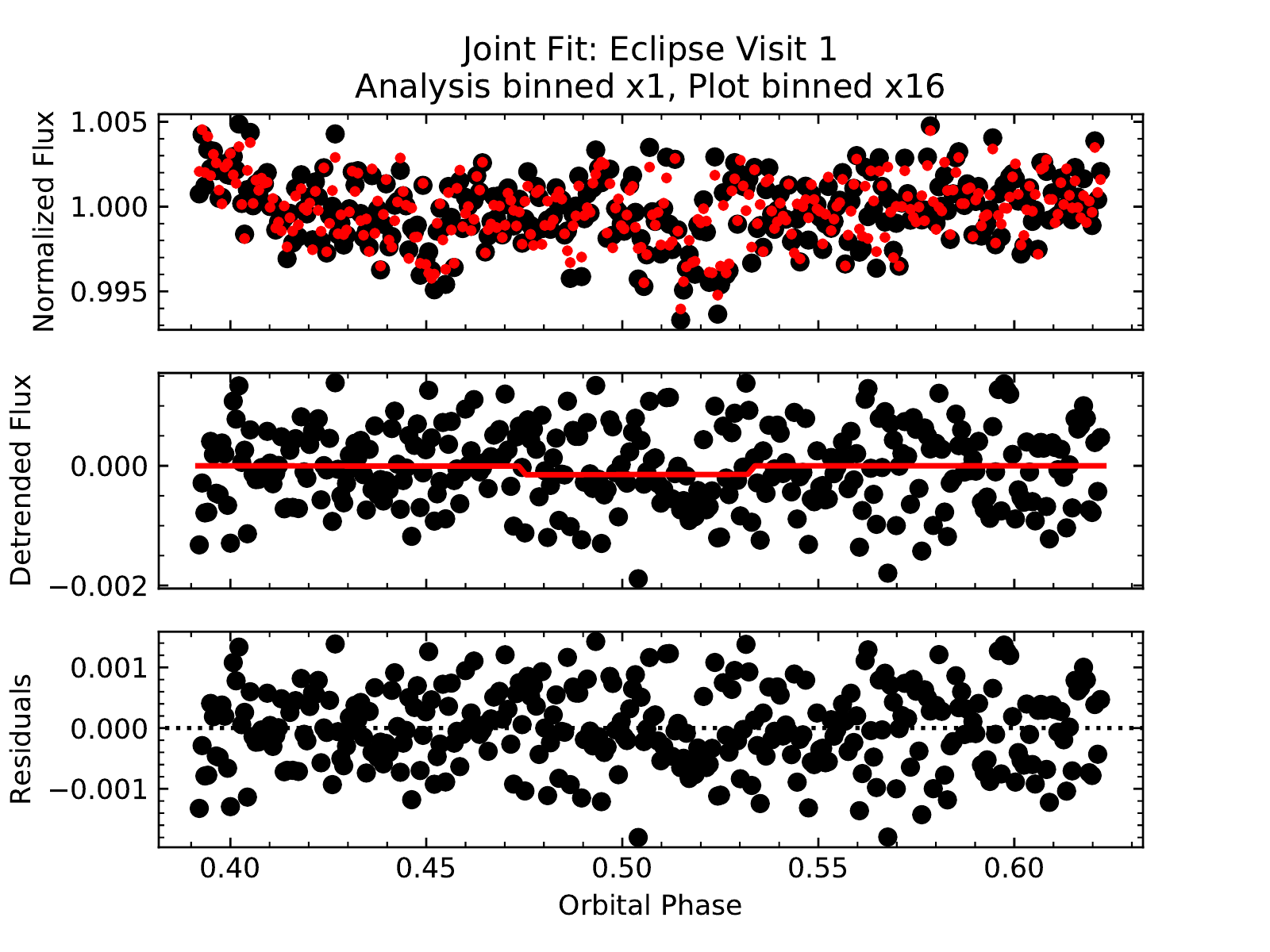}
\includegraphics[width=0.45\textwidth,page=2]{20220407_single-eclipse_plots}
\includegraphics[width=0.45\textwidth,page=3]{20220407_single-eclipse_plots}
\includegraphics[width=0.45\textwidth,page=4]{20220407_single-eclipse_plots}
\includegraphics[width=0.45\textwidth,page=5]{20220407_single-eclipse_plots}
\includegraphics[width=0.45\textwidth,page=6]{20220407_single-eclipse_plots}
\caption{{\em Spitzer} 4.5\,\micron\ light curves for our joint
  analysis of all eclipse visits.  In each triptych
  the top panel shows the raw {\em Spitzer} photometry (black circles)
  vs.\ the PLD model (red points); the middle panel shows the
  systematics-corrected photometry (black circles) vs. the best-fit
  eclipse model (red line); the bottom panel shows the
  residuals. Although we performed our analysis without binning the
  data, the data are binned here by a factor of 16 for plotting
  purposes. Fig.~\ref{fig:eclipse} shows the stacked average of all ten eclipses.
  \label{fig:singlevisits1}}
\end{figure*}

\begin{figure*}
\centering
\includegraphics[width=0.45\textwidth,page=7]{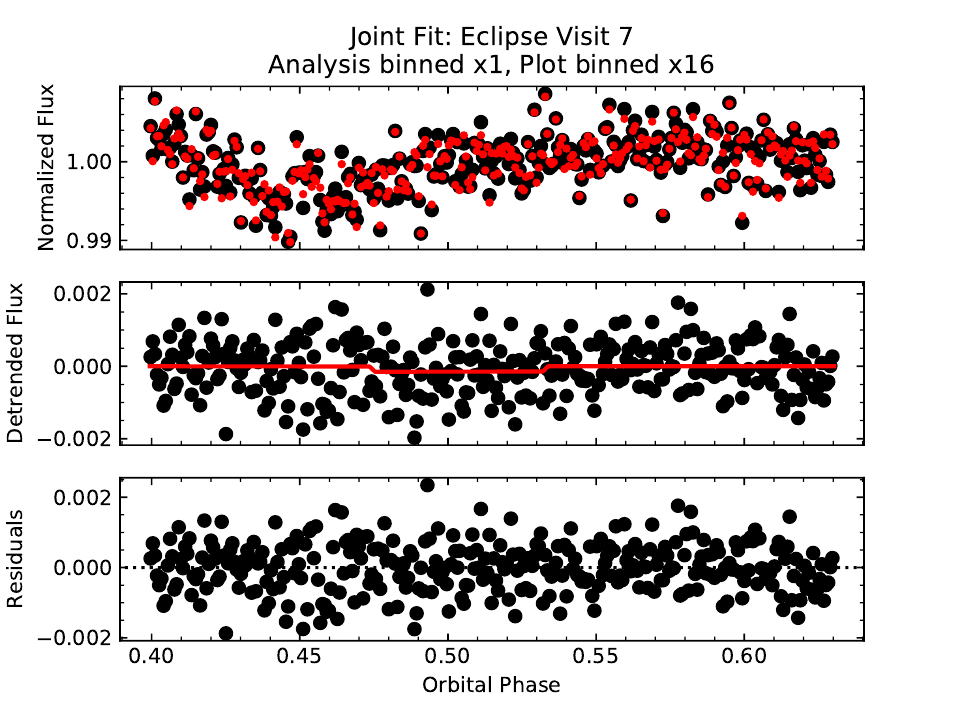}
\includegraphics[width=0.45\textwidth,page=8]{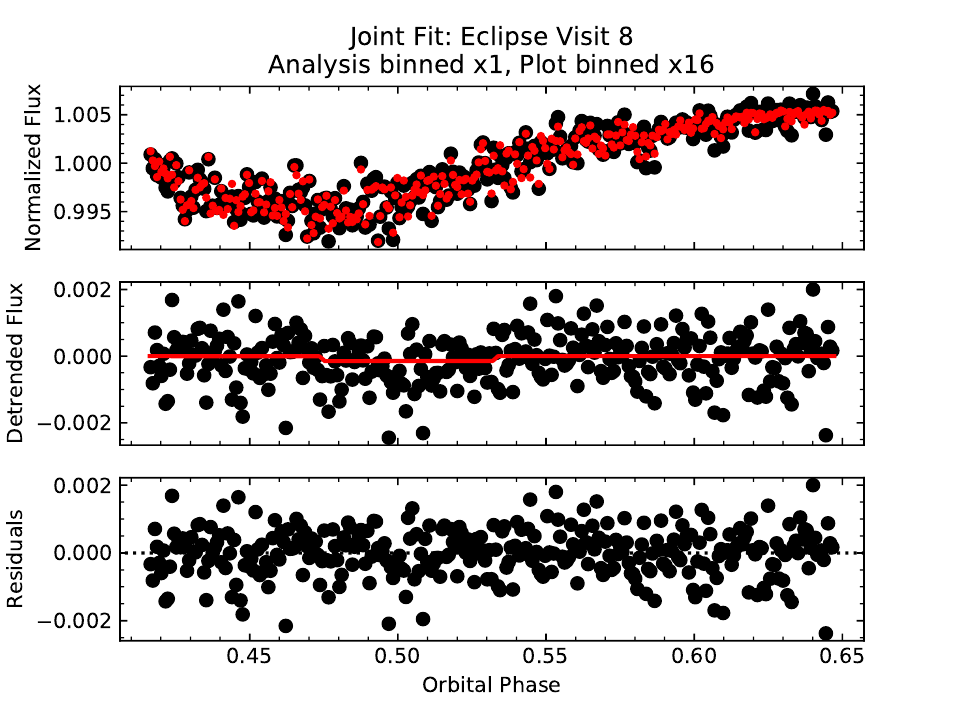}
\includegraphics[width=0.45\textwidth,page=9]{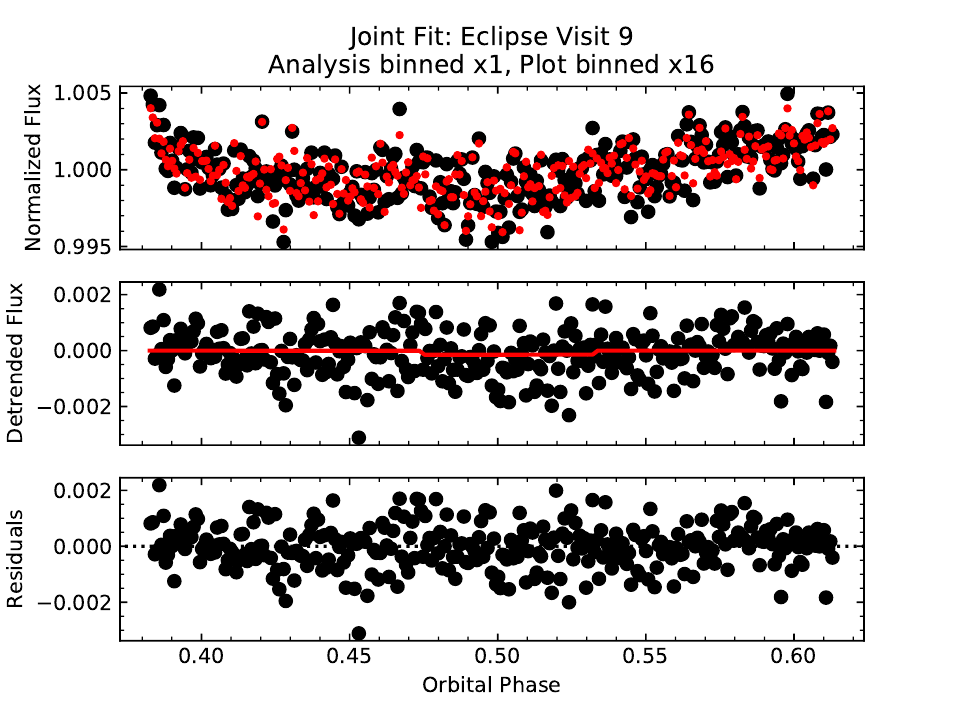}
\includegraphics[width=0.45\textwidth,page=10]{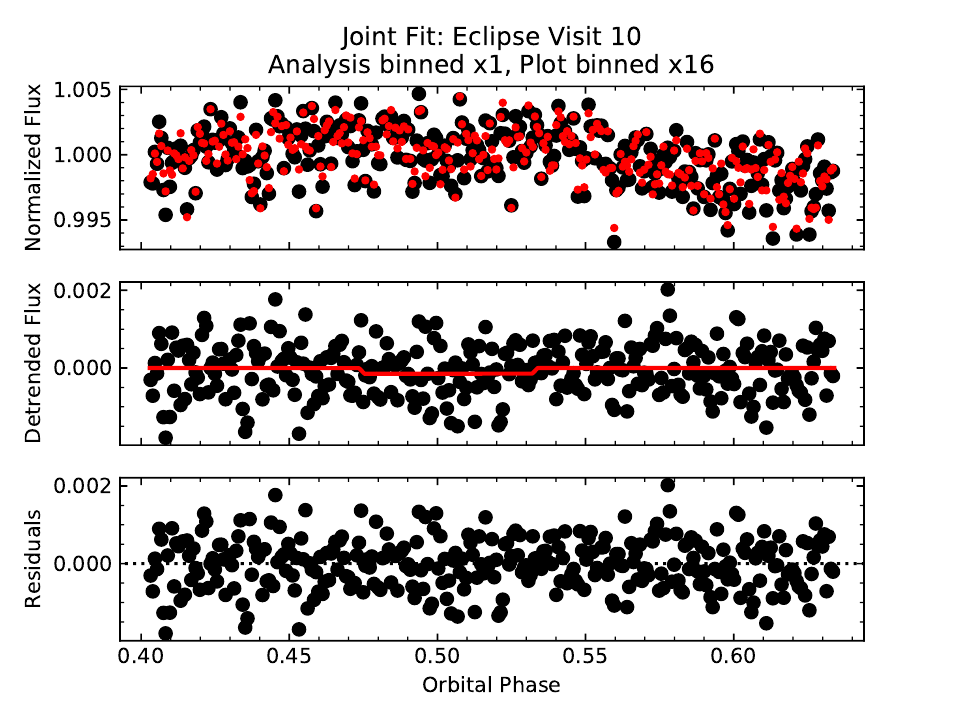}
\caption{Same as Fig.~\ref{fig:singlevisits1}, but for the final four {\em Spitzer} eclipse visits.
  \label{fig:singlevisits2}}
\end{figure*}

\begin{figure*}
\centering
\includegraphics[width=\textwidth]{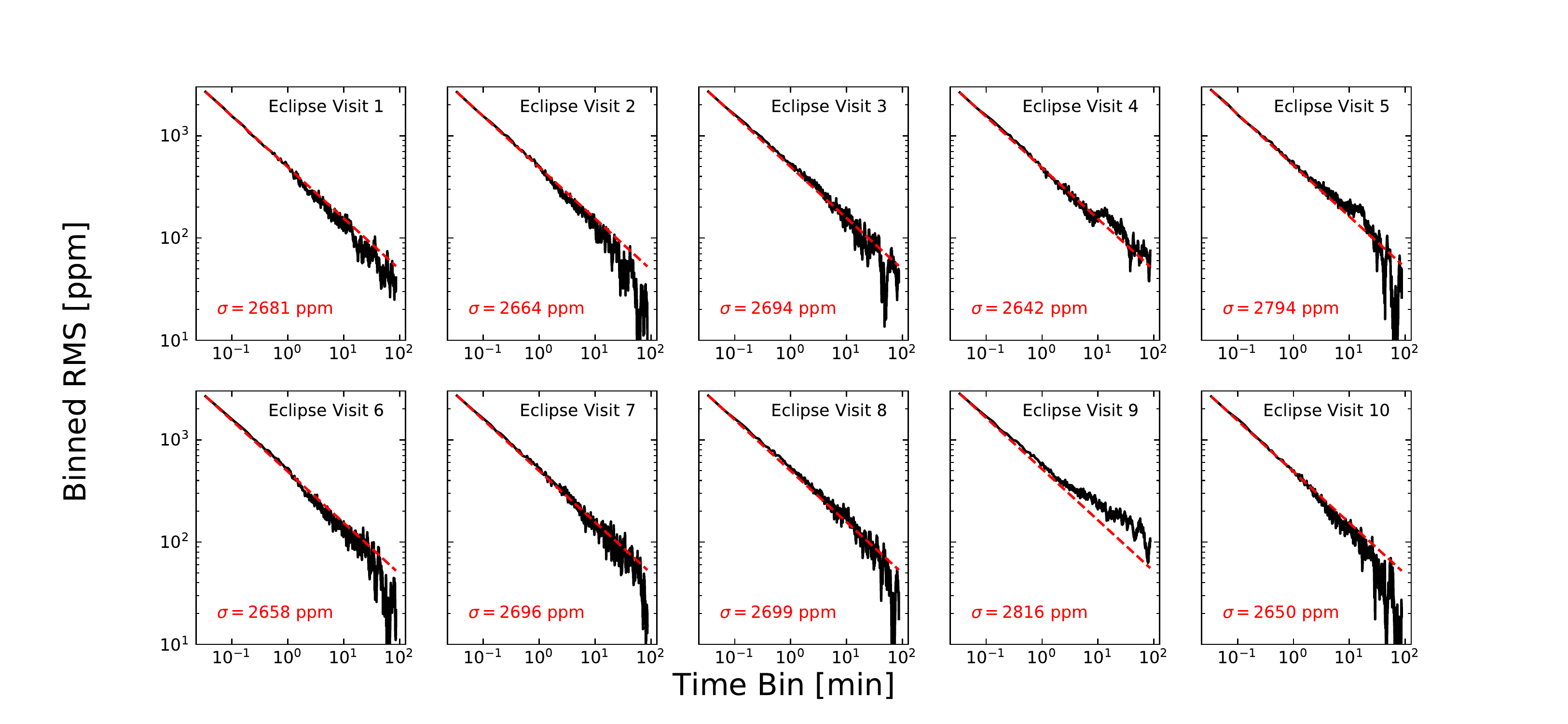}
\caption{The dispersion of the binned residuals (solid black line) to
  the 4.5\,\micron\ light curves show limited evidence for correlated
  noise. The dashed line shows the expectation for wholly uncorrelated
  errors, which scale as $N^{-1/2}$.
  \label{fig:bindown}}
\end{figure*}

Fig.~\ref{fig:corner} shows the joint posteriors on eclipse depth and
time of eclipse for our primary, joint
analysis. Figs.~\ref{fig:singlevisits1} and~\ref{fig:singlevisits2}
show the raw {\em Spitzer} photometry for each individual eclipse
visit, the photometry after removal of systematics, and the residuals
to the fits.  Fig.~\ref{fig:bindown} shows how the residuals to these
individual fits bin down with time.

\end{document}